\documentclass[preprint,cleveref=noabbrev]{elsarticle} 

\makeatletter
\newcounter{ALC@unique}
\makeatother

\usepackage[draft=true, cite=false, 
enumitem=true, clevethm=true, builddir=., tikzoutdir=Pics/Tikz, linkcolor=NavyBlue,
mparhack=false, microtype=false]{defaultpackages}
\usepackage{mynotation}
\usepackage{scenariotree}
\usepackage{automotive_notation}
\usepackage{booktabs}

\usetikzlibrary{patterns,positioning,arrows,calc}
\colorlet{blue}{NavyBlue}
\tikzset{main node/.style={circle,fill=blue!20,draw,minimum size=1cm,inner sep=0pt},
}

\newcommand{\revision}[2]{#2}
\newcounter{todo}
\newcommand{\todo}[1][]{\stepcounter{todo}\textcolor{red}{TODO \thetodo\ifstrempty{#1}{}{: #1}}}

\newcounter{experiment}[section]
  \newenvironment{experiment}[1][]{\refstepcounter{experiment}
  \par\medskip
  \noindent\hspace*{-1em}
  \ifstrempty{#1}{\textbf{Experiment~\theexperiment.}}{\textbf{Experiment~\theexperiment~--- #1.}} \leftskip=1em\rmfamily}{
  \medskip
  }
  \numberwithin{experiment}{section}

  \creflabelformat{experiment}{#2#1#3}
  \crefname{experiment}{Experiment}{Experiments}
\usepackage{textcomp}
\usepackage{subcaption}
\usepackage{tabularx}

\newcommand{\IPOPT}{\textsc{ipopt}}
\newcommand{\CASADI}{CasADI}
\newcommand\thetitle{Safe, Learning-Based MPC for 
Highway Driving under Lane-Change Uncertainty: A Distributionally Robust Approach}

\tikzexternaldisable
\begin{document}
\begin{frontmatter}

\title{\thetitle\tnoteref{ack,vid}}
\tnotetext[ack]{This work was supported by the Ford KU Leuven Research Alliance Project (KUL00024, KUL0075). The work of P. Patrinos was supported by KU Leuven internal funding: C14/18/068; Fonds de la Recherche Scientifique (FNRS) and Fonds Wetenschappelijk Onderzoek (FWO) Vlaanderen under EOS Project no G0F6718N (SeLMA); FWO: G0A0920N, G086318N, G086518N; European Union's Horizon 2020 research and innovation programme under the Marie Skłodowska-Curie grant agreement No. 953348.}
\tnotetext[vid]{Supplementary videos are available online \cite{project_page}.}

\author[kul]{Mathijs Schuurmans\corref{cor1}}
\ead{mathijs.schuurmans@kuleuven.be}
\cortext[cor1]{Corresponding author}
\author[fordaachen]{Alexander Katriniok}
\author[fordgreenfield]{Christopher Meissen} 
\author[forddearborn]{H. Eric Tseng}
\author[kul]{Panagiotis Patrinos}

\affiliation[kul]{organization={Department of Electrical Engineering (ESAT-STADIUS), KU Leuven},
addressline={Kasteelpark Arenberg 10}, 
city={Leuven},
postcode={3001}, 
country={Belgium}}

\affiliation[fordaachen]{organization={Ford Research \& Innovation Center},
city={Aachen},
postcode={52072}, 
country={Germany}}

\affiliation[forddearborn]{organization={Ford Research \& Innovation Center},
city={Dearborn},
postcode={48124}, 
state={Michigan},
country={USA}}

\affiliation[fordgreenfield]{organization={Ford Greenfield Labs},
city={Palo Alto},
postcode={52072}, 
state={California},
country={USA}}

\begin{abstract}
We present a case study applying
learning-based distributionally robust model predictive control 
to highway motion planning under stochastic uncertainty of 
the lane change behavior of surrounding road users. 
The dynamics of road users are modelled using Markov jump 
systems, in which the switching variable describes the 
desired lane of the vehicle under consideration 
\revision{}{and the continuous state describes the pose and velocity of the vehicles}.
We assume the switching probabilities of the underlying Markov 
chain to be unknown. As the vehicle is observed and thus, samples 
from the Markov chain are drawn, the transition 
probabilities are estimated along with an ambiguity set which 
accounts for misestimations of these probabilities. Correspondingly, 
a distributionally robust optimal control problem is formulated 
over a scenario tree, and solved in receding horizon. As 
a result, a motion planning procedure is obtained which
through observation of the target vehicle gradually becomes less 
conservative while avoiding overconfidence in estimates obtained from 
small sample sizes. We present an extensive numerical 
case study, comparing the effects of several different design aspects 
on the controller performance and safety. 
\end{abstract}

\begin{keyword}
Model predictive control \sep Risk measures \sep Distributionally robust optimization \sep Automated driving \sep Path planning

\MSC[2010] 93E20\sep93C10\sep93E35\sep91G70
\end{keyword}

\end{frontmatter}
\section{Introduction} \label{sec:intro}

The prospect of significantly increasing safety and efficiency 
of traffic have motivated considerable research efforts towards increasingly advanced driving assistance systems and even (fully) autonomous driving. 
Due to their strong theoretical foundations and the ability 
to naturally include safety constraint\revision{}{s} in the control 
design, systems based on \ac{MPC} provide promising avenues towards 
this goal.
However, one of the key challenges in planning and following safe trajectories among neighboring traffic participants is the 
large amounts of uncertainty present in forecasts of the traffic state in the near future \cite{carvalho_AutomatedDrivingRole_2015a}.

This central challenge has led to the growing popularity of stochastic 
\ac{MPC} formulations using models obtained through techniques 
from \ac{ML} and statistics \cite{lefevre_SurveyMotionPrediction_2014a}.
However, among all domains where \ac{ML} techniques provide great promise, it is clear that autonomous driving is one where interpretability, safety and reliability of the employed models are of vital importance. 
This is a concern with many commonly used \ac{ML} models,
in particular at smaller sample sizes.
\revision{}{Indeed, it is well-known that sample-average approximations for minimizing the expected value of random cost functions under unknown 
distributions often leads to poor out-of-sample performance. This approach, which is closely related to overfitting in machine learning, is known in the optimization and operations research community as the optimizer's curse \cite{smith_OptimizerCurseSkepticism_2006a,parys_DataDecisionsDistributionally_2020}.}

Motivated by these concerns, robust approaches are traditionally 
considered as a safe alternative \cite{lofberg_ApproximationsClosedloopMinimax_2003,batkovic_RobustScenarioMPC_2021}. Here, one would replace a probability 
distribution over realizations of an uncertain influence on the system 
by the worst-case realization over a prediction horizon. As such, 
only the knowledge of the support of the disturbance distribution 
(i.e., the uncertainty set) is required to provide \emph{a priori} safety guarantees.
The obvious drawback of such formulations is that, in practice, requiring
robustness to all possible realization may be too strict.
The resulting control formulation may be overly conservative, and therefore lead to severe loss of driving efficiency or infeasibility 
(situations in which no action or policy satisfying the safety constraints exists). Furthermore,
empirical evidence suggests that overly cautious behavior may even reduce safety \cite{schoettle_PreliminaryAnalysisRealWorld_2015}.

In order to capture the benefits of both approaches,
while mitigating their respective drawbacks,
we propose a distributionally robust control methodology \cite{rahimian_DistributionallyRobustOptimization_2019,dupacova_MinimaxApproachStochastic_1987}.
To do so, we consider a stochastic model of the joint dynamics of the 
agents in the immediate traffic environment assuming that the underlying distribution is unknown.
In every time step, the proposed controller receives a 
sample from this distribution through observing the current state of the system. Using these gathered samples, the parameters of a stochastic 
model are estimated, along with statistical bounds on the possible misestimation of these parameters. Combining these two quantities 
yields a set of statistically plausible distributions over the 
stochastic process in the next time step---these are commonly referred to as \emph{ambiguity sets}.

The controller then plans a trajectory assuming the worst-case probability distribution within this ambiguity set, resulting in a formulation equivalent to a risk-averse \ac{OCP} \cite{sopasakis_RiskaverseModelPredictive_2019,sopasakis_RiskaverseRiskconstrainedOptimal_2019}. Initially, all ambiguity sets are expected to contain (almost) all distributions (leading to a robust \ac{OCP}), whereas over time, the uncertainty on the estimated distributions decreases and accordingly, the conservatism of the corresponding control decisions is systematically reduced. 

In this work, we focus on the task of highway path planning,
considering stochastic behavior of the surrounding vehicles (target vehicles).
We model the target vehicle states as autonomous 
Markov jump systems; 
At every time step, the driver of the target vehicle makes a choice of semantically meaningful driving behaviors, which we refer to as the \emph{mode} of the system. In particular, each mode could correspond to a desired lane of the vehicle. Given the (randomly) selected mode, the 
continuous state of the vehicle is updated using a kinematic vehicle model. Repeating this over multiple time-steps yields a \revision{}{stochastic, yet} dynamically feasible state trajectory.

The choice of this model class is motivated by the typical multimodal nature of distributions of vehicle trajectories on highways \cite{phan-minh_CoverNetMultimodalBehavior_2020}.
Indeed, the majority of lateral position trajectories will be concentrated around the lane center lines, whereas only a small minority of trajectories will run along the separating lines between the lanes.
\revision{
By contrast, a Gaussian model, for instance, would typically assign the largest probability to the latter, as it will approximately correspond to the mean trajectory.
}{
By contrast, many popular models for representing stochasticity in control systems, such as Gaussian processes, will fail to represent this bimodality.
}

Furthermore, under this model, the resulting optimal control problem can be represented on a scenario tree,
which allows us to optimize over all causal closed-loop policies without requiring a parametric approximation,
reducing conservatism with respect to alternative approaches,
such as tube-based formulations or affine state-feedback parametrization \cite{bonzanini_SafeLearningbasedModel_2020,summers_DistributionallyRobustSamplingBased_2018}.

\subsection{Related work}
Within the field of automated driving and advanced driver assistance systems, the same class of switching models was used in earlier work \cite{schuurmans_learningBased_2020a,bichi_stochastic_2010}
for \iac{ACC} setting.
In \cite{bichi_stochastic_2010}, this model is employed in a stochastic 
\ac{MPC} framework. In \cite{schuurmans_learningBased_2020a}, this 
was extended to a distributionally robust setting, where 
the transition probabilities were no longer required to be known \emph{a priori}. In this work, we subsequently extend the work of \cite{schuurmans_learningBased_2020a} to a full path planning use case, which includes both longitudinal and lateral control as well as possibly nonlinear vehicle dynamics and (inherently nonconvex) collision avoidance constraints.

To ensure safety of the generated trajectories, we impose collision avoidance constraints in the form of conditional \emph{chance constraints}. For numerical reasons, however, we replace chance constraints by risk constraints
\cite[\S 3.1.1]{nemirovski_SafeTractableApproximations_2012}. 
Risk measures been used for collision avoidance in the broader context of autonomous navigation under uncertainty
\revision{
  \cite{
  hakobyan_risk-aware_2019,
  chapman_risk-sensitive_2019,
  dixit_RiskSensitiveMotionPlanning_2021,
  ahmadi_RiskAversePlanningUncertainty_2020,
  ono_ChanceconstrainedDynamicProgramming_2015,
  hakobyan_ImprovingDistributionalRobustness_2021,
  summers_DistributionallyRobustSamplingBased_2018
}.   
}{
\cite{
hakobyan_risk-aware_2019,
chapman_risk-sensitive_2019,
dixit_RiskSensitiveMotionPlanning_2021,
ahmadi_RiskAversePlanningUncertainty_2020,
ono_ChanceconstrainedDynamicProgramming_2015,
hakobyan_ImprovingDistributionalRobustness_2021,
nishimura_RiskSensitiveSequentialAction_2020,
samuelson_SafetyAwareOptimalControl_2018,
summers_DistributionallyRobustSamplingBased_2018,
chen_InteractiveMultiModalMotion_2022}.
}

In \cite{
chapman_risk-sensitive_2019,
dixit_RiskSensitiveMotionPlanning_2021,
ahmadi_RiskAversePlanningUncertainty_2020,
nishimura_RiskSensitiveSequentialAction_2020,
samuelson_SafetyAwareOptimalControl_2018,
chen_InteractiveMultiModalMotion_2022}, risk measures are constructed with respect to some known reference distribution,
treating the risk level as a tuning parameter.
In contrast to these approaches, we adopt a \textit{distributionally robust} approach, in which risk measures are implicitly defined through statistical ambiguity sets
(see \cref{sec:methodology} for more details).
As a result, risk parameters are automatically calibrated based on the sample size of the observed dataset \cite{schuurmans_GeneralFrameworkLearningBased_2021}.

This distributionally robust perspective is shared with
\cite{hakobyan_ImprovingDistributionalRobustness_2021,summers_DistributionallyRobustSamplingBased_2018,hakobyan_WassersteinDistributionallyRobust_2020,wang_NonGaussianChanceConstrainedTrajectory_2020},
which approximate chance constraints using concentration inequalities. Although these approaches are similar in spirit, they differ significantly in the remaining modeling choices and solution methodologies.
For instance, contrary to our scenario tree formulation, which models discrete probabilistic choices, the cited works consider 
the uncertainty to originate from additive disturbances \cite{summers_DistributionallyRobustSamplingBased_2018,hakobyan_wasserstein_2020}, or from the uncertainty of a Gaussian process model of the environment dynamics \cite{hakobyan_ImprovingDistributionalRobustness_2021}. These modeling choices in turn dictate the use of different classes of ambiguity sets (e.g., moment-based ambiguity sets \cite{summers_DistributionallyRobustSamplingBased_2018}, Wasserstein balls \cite{hakobyan_ImprovingDistributionalRobustness_2021}), 
and corresponding optimal control formulations and solution methods. 

On the other hand, in the model of \cite{wang_NonGaussianChanceConstrainedTrajectory_2020}, qualitative (discrete) driver actions can be taken into account,
as they allow multimodal mixture models to describe the distributions of \revision{future the}{future} target vehicle states.
\revision{Here, however}{However, in contrast to our work}, it is assumed that the distributions are known (i.e., provided by some given learning system).
The purpose of the concentration inequalities in \cite{wang_NonGaussianChanceConstrainedTrajectory_2020} -- which depend on these distributions -- is therefore to ease computations, rather than to ensure robustness against ambiguity.
\revision{}{
Furthermore, several techniques to predict future trajectories of surrounding vehicles, based on 
continuous/discrete hybrid models, e.g., based on Hidden Markov Models \cite{liu_DrivingIntentionPrediction_2019} 
or Long Short-Term Memory (LSTM) models \cite{deo_MultiModalTrajectoryPrediction_2018,huang_HYPERLearnedHybrid_2021} have been proposed in 
the literature. 
However, it is very difficult to derive practically useful and statistically valid concentration bounds on the obtained probability distributions
for the purpose of constructing ambiguity sets.
Furthermore, making predictions of future states using e.g., the method of \cite{deo_MultiModalTrajectoryPrediction_2018}
requires performing inference on the LSTM model. 
By consequence, trajectory prediction and planning must typically be carried out separately, since 
otherwise, solving the optimal control problem involves running inference of the LSTM, including its gradient at every iteration. 
By contrast, with the model class used in this work, it is easy to introduce dependence of the target vehicle policy on the ego vehicle states,
which affords additional modeling freedom for the purpose of integration into \iac{MPC} scheme. We provide an example of this in \Cref{sec:multi-vehicle-experiment}.
}

\subsection{Contributions and organization}

We now summarize the main contributions of this work: 
\begin{enumerate}
 \item We propose a simple, interpretable and extendable model for the behavior of target vehicles in highway scenarios. This model takes the form of a Markov jump system, which naturally models both the 
 uncertainty inherent to the tactical driving decisions of the driver (represented by the discrete modes) and the more predictable nature of vehicle kinematics. 
 \revision{}{Compared to existing, more sophisticated discrete/continuous hybrid models from the literature
 (e.g., \cite{deo_MultiModalTrajectoryPrediction_2018,huang_HYPERLearnedHybrid_2021}),
 these have the benefit of being easy to encorporate into an MPC scheme, and allowing for dependence of their dynamics on the ego vehicle state.}
\item We present a distributionally robust \ac{MPC} scheme for closed-loop path planning under probabilistic collision avoidance constraints. The true underlying distributions are assumed to be unknown. As the target vehicle behavior is observed online, the nominal transition probabilities are estimated, along with an \emph{ambiguity set} which models the
statistical uncertainty on the estimated distribution.
\item We carry out an extensive numerical case study, illustrating the effects of the involved design parameters on the behavior of the control 
system. This includes a comparison of several formulations of the collision avoidance constraints. 
\end{enumerate}

The remainder of the article is organized as follows: 
In \Cref{sec:modelling}, we describe the set-up of the case study. We describe the used vehicle models, and introduce several possible formulations describing collision-avoidance constraints. In \Cref{sec:methodology}, we briefly review some technical preliminaries and introduce the \ac{MPC} design. 
\Cref{sec:case-study} presents an extensive numerical case study of the resulting control scheme. Finally, in \Cref{sec:conclusion}, we discuss the most salient 
conclusions and suggest possible extensions and improvements for 
future work.

\section*{Notation}
Let $\Re$ and $\N$ denote the set of real and natural numbers, respectively. $\Re_+$ ($\Re_-$) denotes the set of positive (negative) real numbers. $I_n$ denotes the $n\times n$ identity matrix. 
Given $a, b \in \N, a \leq b$, we write $\natseq{a}{b} \dfn \{\revision{k}{n} \in \N \mid a \leq n \leq b \}$. 
We use $\1_\nModes$ to denote a vector in $\Re^\nModes$ with all elements equal to 1.
For some $i \in \natseq{1}{\nModes}$, we denote by  $\e_{i} \in \Re^{\nModes}$ the $i$'th standard basis vector,
taking the value 1 in its $i$'th components and 0 everywhere else. Given a matrix $P$,
we denote its $i$th row by $\row{P}{i}$.
We denote by $[x]_{+}$ the positive part of $x$, i.e., $[x]_{+} = \max\{0, x\}$, where $\max$ is evaluated element-wise.

\section{Modelling for highway motion planning} \label{sec:modelling}
The following section describes the set-up and the models used in 
our case study. This involves the dynamics of the host \revision{}{(or ego)} vehicle,
of \emph{target vehicles} (i.e., surrounding vehicles)
and the considered formulations for the obstacle avoidance constraints.
Overall, we will consider driving scenarios on a multi-lane highway.
For simplicity, we consider straight road segments. As our final controller does not rely on 
convexity or linearity of the involved dynamics, the set-up can be straightforwardly extended 
to more general curved roads by transforming all involved coordinates 
to a Fren\'{e}t frame \cite{werling_OptimalTrajectoryGeneration_2010}.

We will denote the state of the $i$'th vehicle in the scene by $\id{x}{i}$. We will assume the convention that vehicle 0 is the ego vehicle. The augmented state vector obtained by combining 
the states of both the ego vehicle and the target vehicles will be 
denoted as $\xb = \trans{\smallmat{\id{x}{0} & \dots & \id{x}{n}}} \in \Re^{\ns}$. 
\revision{}{This allows us to compactly write the dynamics of the full concatenated system as
\begin{equation} \label{eq:dynamics-general}
   \xb_{t+1} = f(\xb_t, u_t, \md_{t+1}),
\end{equation}
where $u_t \in \Re^{\na}$ denotes the control actions applied to the ego vehicle at time $t$ and $\{\md_t\}_{t \in \N}$
is a stochastic process, representing the uncertain behavior of the surrounding vehicles.
We will assume that $\md_t$ takes values on a finite set $\W \dfn \{1, \dots, \nModes \}$ 
and is governed by a time-homogeneous Markov chain with \textit{unknown} transition probability matrix $\transmat$.
In \Cref{sec:learning}, we describe how we use online observations to obtain estimates 
of $\transmat$, while remaining aware of potential estimation errors.
In the next sections (\cref{sec:dynamics-ego,sec:dynamics-tv}), 
we provide example implementations for $f$, which we have used for our numerical experiments. 
However, we do emphasize that many other modeling choices for $f$ could be made without 
any modification to the remainder of the control formulation, which is described in \Cref{sec:methodology}.
}

\subsection{Host vehicle Dynamics} \label{sec:dynamics-ego}
The host vehicle dynamics are considered to be deterministic and known.
\revision{
We will perform experiments using a simplified linear model as well as a more realistic nonlinear model.
In the first experiments (\cref{exp:experiment-1}), the host vehicle dynamics are modelled as decoupled double integrator
systems in the longitudinal and lateral directions. 
We will refer to this simplified model as the \textit{double integrator model}, 
the dynamics of which is given by 
\begin{equation} \label{eq:double-integrator-ego}
    x_{t+1} = \smallmat{1 &   & \Ts &  \\ 
                          & 1 &     & \Ts\\
                          &   &  1  &   \\ 
                          &   &     & 1} x_t 
                + 
              \smallmat{ 0_{2 \times 2} \\ 
                         \Ts I_2} u_t. 
\end{equation}
\todo[Mention in the rebuttal that we removed the linear model.]
Dynamics \eqref{eq:double-integrator-ego} describes the evolution of the state vector $x = \trans{\smallmat{\pos_x &\pos_y & \vel_x & \vel_y}}$, which represent the positions and velocities in the longitudinal ($x$) and lateral ($y$) direction with respect to the inertial road frame. The control actions $u = \trans{\smallmat{\acc_x & \acc_y}}$ describe the longitudinal and lateral accelerations of the vehicle.
In this model, the longitudinal and lateral motion are completely 
decoupled, which, under a no-slip assumption on the tires, 
allows for physically inconsistent trajectories. Although in practice, we have not found this to be of major consequence as this coupling 
can be reintroduced by defining a small constant $c>0$ and imposing a constraint $-c \vel_x \leq \vel_y \leq c \vel_x$ as proposed in \cite{nilsson_ManoeuvreGenerationControl_2014}, for instance.
Later experiments use a more realistic and commonly used dynamics model known
as the
}
{We utilize the well-known}
 kinematic bicycle model \cite{rajamani_VehicleDynamicsControl_2012},
which is governed by continuous-time dynamics of the form 
    \begin{subequations}
        \label{eq:bicycle-model-ego}
        \begin{align} 
            \dot{\pos_x} &= \vel \cos(\psi + \beta) \label{eq:bicycle-model-pos-x} \\ 
            \dot{\pos_y} &= \vel \sin(\psi + \beta) \label{eq:bicycle-model-pos-y}\\
            \dot{\psi}   &= \tfrac{\vel}{\lr} \sin(\beta) \label{eq:bicycle-model-angle}\\ 
            \dot{\vel}   &= a \label{eq:bicycle-model-vel}\\
            \beta &= \tan^{-1} \left( \tfrac{\lr}{\lf + \lr} \tan{\delta} \right) \label{eq:bicycle-model-beta}
        \end{align}
    \end{subequations}
with state vector $x = \trans{\smallmat{\pos_x & \pos_y & & \psi & \vel}}$
(longitudinal and lateral position, heading angle and longitudinal velocity)
and input vector $u = \trans{\smallmat{a & \delta}}$ (longitudinal acceleration 
and steering angle).
In our experiments, we discretize these equations using a simple forward Euler scheme, although of course, more sophisticated integration schemes could readily be used for our control scheme, if additional accuracy were required. For simplicity, we will set $\lr = \lf = \tfrac{l}{2}$.

\subsection{Target vehicle behavior} \label{sec:dynamics-tv}
\revision{The target vehicle dynamics are given by an autonomous \ac{MJLS} \cite{costa_DiscretetimeMarkovJump_2005} which has a separate driving mode corresponding to each lane on the road.
}{
We model the target vehicle behavior as an autonomous Markov jump system, which includes a separate driving mode corresponding to each lane on the road.
}
Within each mode, 
the dynamics are governed by an autonomous system that drives the lateral component of the vehicle position to the centerline of the lane corresponding to the current mode
and the longitudinal component to a preferred velocity (given for each lane).
\revision{}{ Additionally, it is possible for the policy to take into account relative positions and velocities with respect to other vehicles.}
The state vector $\id{x}{i} = \trans{\smallmat{\id{\pos_x}{i} & \id{\pos_y}{i} & \id{\vel_x}{i} & \id{\vel_y}{i}}}$ of vehicle $i$ consists of the position and velocity in the $(x,y)$-plane.
For simplicity, the longitudinal and lateral dynamics are decoupled for these vehicles into two double-integrator systems,
leading to dynamics of the form
\revision{
$x_{t+1} = A x_t + p(\md_{t+1})$, with
\begin{equation} \label{eq:target_vehicle_dyn_old}
\begin{aligned}
        &A = \smallmat{
                    1  &0         & \Ts              & 0                  \\ 
                    0  &1         &  0               & \Ts               \\
                    0  &0         &  1 - \Ts \kvx & 0                 \\
                    0  & -\Ts \ky &  0               & 1 - \Ts \kvy 
                    },
        \\
        &p(\md) = \smallmat{0_{2\times 2}\\ \Ts I_2} 
               \smallmat{
                         \kvx \sub{{\vref}}{,\md} \\ 
                         \ky \sub{y_\textrm{lane}}{\md}
                        },
\end{aligned}
\end{equation}
}{
\begin{equation} \label{eq:target_vehicle_dyn}
\begin{aligned}
    \id{x}{i}_{t+1} = A \id{x_t}{i} + B \id{\pi}{i}\left( \xb_t, \md_{t+1} \right), \text{ with }
    A = \smallmat{
        1  &\Ts       &  0               & 0      \\ 
        0  &1         &  0               & 0      \\
        0  &0         &  1               & \Ts    \\
        0  &0         &  0               & 1 
        }
    ,\quad 
    B = \smallmat{0 & 0\\ 
                 \Ts & 0\\
                 0  & 0\\
                 0  & \Ts},
\end{aligned}
\end{equation}
and $\id{\pi}{i}: \Re^{\ns} \times \W \to \Re^{2}$ is the mode-dependent control policy for target vehicle $i$. 
Unless otherwise specified, we will assume that the policy is given as, 
\begin{equation} \label{eq:policy-tv-simple}
    \id{\pi}{i}(\xb, \md) = \smallmat{
        0  & 0     & \kvx & 0\\ 
        0  & -\ky  &  0    & -\kvy
    } \id{x}{i} + 
    \smallmat{
        \kvx \sub{{\vref}}{,\md} \\ 
        \ky \sub{y_\textrm{lane}}{\md}
    }.
\end{equation}
See \Cref{tab:symbols} for a description of the involved parameters.
Note that under \eqref{eq:policy-tv-simple}, the target vehicle dynamics 
are described as an autonomous \ac{MJLS} \cite{costa_DiscretetimeMarkovJump_2005}. 
However, we emphasize that the presented control framework allows for sophisticated models for the target vehicle behavior, as we illustrate in \cref{sec:multi-vehicle-experiment}.
}

\revision{
where we have dropped the vehicle index $i$ for ease of notation. 
Here, $(\md_t)_{t \in \N}$ is a Markov chain with (unknown) transition kernel $\transmat$, taking values on the finite set $\W \dfn \{1, \dots, \nModes\}$.
See \Cref{tab:symbols} for a description of the involved parameters. 
}{}

\revision{}{The decoupling of the nonlinear vehicle dynamics into longitudinal and lateral motion
 is a common modeling choice for highway driving applications
(e.g., \cite{leurent_RobustAdaptiveControlLinear_2020,nilsson_LaneChangeManeuvers_2017,nilsson_ManoeuvreGenerationControl_2014}).
It is motivated by the fact that for highway traffic, 
\begin{inlinelist*}
    \item vehicles tend to be closely aligned with the direction of the road, so that relative heading angles 
    are close to zero
    \item velocities tend to vary slowly around the reference speed.
\end{inlinelist*}
Therefore, a linearization of the nonlinear kinematic bicycle model around the desired velocity
and heading angle $\psi = 0$, leading to \eqref{eq:target_vehicle_dyn} is typically quite accurate.
}

We assume here that the parameters involved in \eqref{eq:target_vehicle_dyn} 
are given. In practice, data-driven estimates of these parameters would need to be obtained.
However, this is beyond the scope of the current paper and considered for future work. 
We assume that at each time $t$, the mode $\md_t$ can be measured, for example by means of the indicator lights of the target vehicle, its 
lateral motion, or a combination of both. See, e.g., \cite{schuurmans_DatadrivenDistributionallyRobust_2021} for a procedure 
to determine mode observations from the continuous state.

\begin{table}[ht!]
    \centering
    \caption{Symbols used in modeling the lane change scenario.}
    \label{tab:symbols}
    \begin{tabularx}{\columnwidth}{@{}lX@{}}
    \toprule
    Parameter & Description                                     \\ \midrule
    $\ky, \kvx, \kvy$             & Internal control gains used to model target vehicle behavior \\
    $\vref$                       & Reference velocity \\
    $y_{\textrm{lane}, i}$        & Lateral position of the lane center of lane $i$ \\
    $\Ts$                          & Sampling rate\\
    $\lr$, $\lf$                  & Longitudinal distance from center of gravity 
                                    and rear/front wheels\\
    $l$, $w$                      & length ($l = l_r + l_f$) and width of a vehicle\\ 
    
    \bottomrule
    \end{tabularx}
\end{table}

\subsection{Collision avoidance formulations} \label{sec:collision-avoidance-forms}

In order to ensure safety of the obtained trajectories, we impose
collision avoidance constraints in the optimal control problem. 
Suppose we are given the augmented state $\xb \in \Re^{\ns}$ of the vehicles in the scene. Then, the obstacle avoidance constraint will 
be of the form 
\begin{equation} \label{eq:obstacle-avoidance-abstract}
    h(\xb) \leq 0, 
\end{equation}
where, in the case of a single pair of vehicles, the function $h: \Re^{\ns} \to \Re$ represents a negative distance between the vehicles. 
Below, we propose several alternatives for the choice of this function
$h$, assuming a single ``obstacle'' vehicle. In the case where more vehicles are present in the scene, the pointwise maximum of the resulting 
functions can be used. It is important to emphasize at this point that 
in the optimal control problem, the future states cannot be exactly predicted. Therefore, a probabilistic counterpart to \eqref{eq:obstacle-avoidance-abstract} is enforced in practice (see \eqref{eq:collision-constraint}). This is discussed in more detail in \Cref{sec:methodology}. For now, however, we focus on the possible instances of 
the function $h$. 

\begin{figure}[ht!]
    \centering
    \newlength{\trimh}
    \setlength{\trimh}{4.5cm}
    \begin{subfigure}[b]{0.45\columnwidth}
        \centering
        \includegraphics[width=\textwidth, trim=0 {\trimh} 0 {\trimh}, clip]{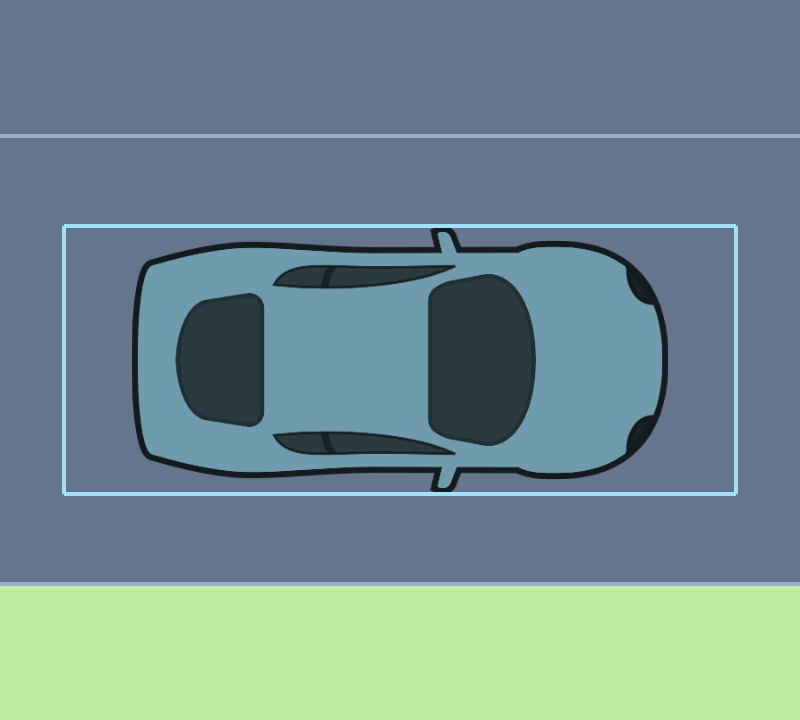}
        \caption{Heading angle $\psi = 0$.}
    \end{subfigure} \hfill
    \begin{subfigure}[b]{0.45\columnwidth}
        \centering
        \includegraphics[width=\textwidth, trim=0 {\trimh} 0 {\trimh}, clip]{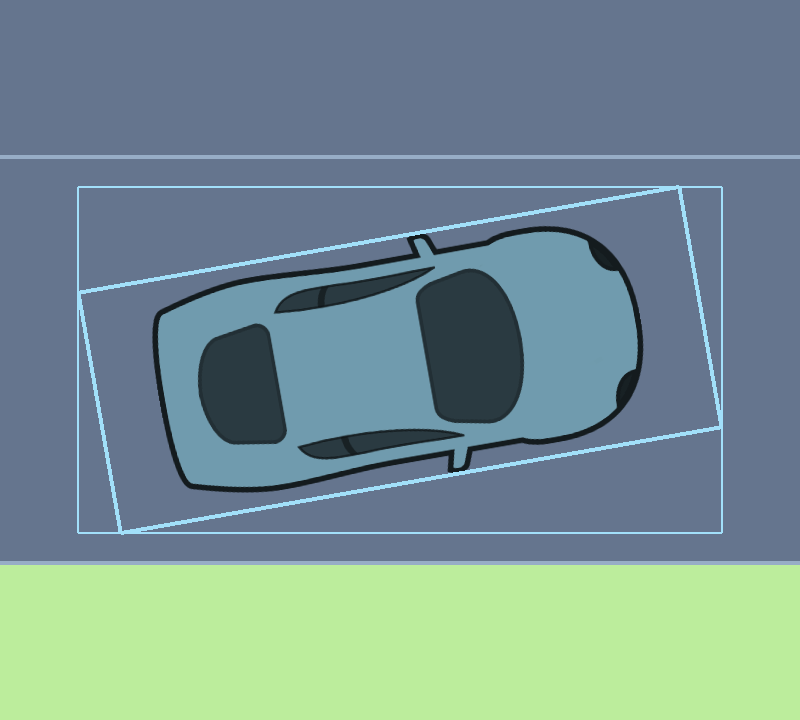}
        \caption{Heading angle $\psi \neq 0$.}
    \end{subfigure}
    \caption{Bounding boxes around a vehicle. For a rotated vehicle, an 
    axis-aligned bounding box induces some additional conservatism.}
    \label{fig:bboxes}
\end{figure}

For the purpose of collision avoidance, we model vehicles using rectangular bounding 
boxes, as illustrated in \cref{fig:bboxes}. The bounding boxes 
can either be aligned with the longitudinal axis of the road or with 
the longitudinal axis of the vehicle. The 
former will typically lead to simplified collision avoidance constraints at 
the cost of some additional conservatism which scales with $|\sin \psi|$.
Since $\sin \psi$ is typically a small value in highway situations, this 
is unlikely to cause significant losses in performance. 

\subsubsection{Axis-aligned bounding boxes}
\paragraph{Computation of axis-aligned bounding boxes}
An axis-aligned rectangle can be represented as the set 
\[ 
    \set{B} = \{\pos \in \Re^2 \mid \pmin \leq \pos \leq \pmax \},
\]
where the inequalities are taken element-wise. If we are given a vehicle with 
length $l$, width $w$ and state $\trans{\smallmat{\pos_x& \pos_y& \psi& v}}$, then $\Delta_y = \tfrac{l}{2} |\sin\psi| + \tfrac{w}{2} |\cos \psi|$,
$\Delta_x = \tfrac{l}{2} |\cos\psi| + \tfrac{w}{2} |\sin \psi|$ and 
\begin{equation} \label{eq:bounding-box}
    \pmax = \smallmat{
        \pos_x + \Delta_x\\
        \pos_y + \Delta_y 
        },\quad  \pmin = \smallmat{
        \pos_x - \Delta_x\\
        \pos_y - \Delta_y 
        }.
\end{equation}

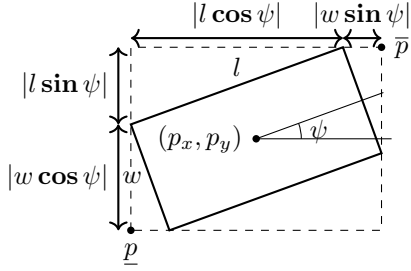
\begin{figure}[htb!]
    \centering
    \begin{tikzpicture}[scale=1]
        \def\xcx{1.2};
        \def\xcy{2.1};
        \def\l{3};
        \def\w{1.5};
        \def\thetadegrees{20}

        \draw[draw=black, rotate=\thetadegrees, thick] (-0.5*\l,-0.5*\w) rectangle (0.5*\l,0.5*\w); 
        \pgfmathsetmacro{\ltot}{(\w * sin(\thetadegrees) + \l * cos(\thetadegrees))}
        \pgfmathsetmacro{\htot}{(\w * cos(\thetadegrees) + \l * sin(\thetadegrees))}
        \coordinate (xmin) at (-0.5*\ltot,-0.5*\htot);
        \coordinate (xmax) at (0.5*\ltot,0.5*\htot); 
        
        \draw[draw=black, dashed] (-0.5*\ltot,-0.5*\htot) rectangle (0.5*\ltot,0.5*\htot); 
        
        \draw[rotate around={\thetadegrees:(0,0)}] (0,0.5*\w) node[above] {$l$};
        \draw[rotate around={\thetadegrees:(0,0)}] (-0.5*\l, 0) node[left,fill=white, fill opacity=0.7, inner sep=0, outer sep=3pt, text opacity=1] {$w$};
        \tikzset{point/.style={circle,  fill=black, inner sep=1}}

        \node[point,  label=below:$\pmin$] at (xmin) {};
        \node[point, label=right:$\pmax$] at (xmax) {};
        \node[point, label=left:{$(\pos_x, \pos_y)$}] at (0,0) {};

        \draw[black, rotate=\thetadegrees] (0,0) -- ({0.6*\l}, 0);
        \draw[black] (0,0) -- ({0.6*\l}, 0);
        \draw ({0.2*\l},0) arc (0:\thetadegrees: 0.2*\l) node[midway, right]{$\psi$} ;

        \draw[<->, black, thick] ({-0.5* (\ltot)}, {0.55*(\htot)}) --++ ({ \l * cos(\thetadegrees))}, 0) node[midway, above]{$|l \cos \psi|$};
        \draw[<->, black, thick] ({0.5* (\ltot)}, {0.55*(\htot)}) --++ ({ -\w * sin(\thetadegrees))}, 0) node[midway, above]{$|w \sin \psi|$};

        \draw[<->, black, thick] ({-0.55* (\ltot)}, {0.5*(\htot)}) --++ (0, {-1*\l * sin(\thetadegrees)}) node[midway, left]{$|l \sin \psi|$};
        \draw[<->, black, thick] ({-0.55* (\ltot)}, {-0.5*(\htot)}) --++ (0, {\w * cos(\thetadegrees)}) node[midway, left]{$|w \cos \psi|$};
    \end{tikzpicture}   
    \caption{Axis-aligned bounding box dimensions}
\end{figure}

Due to the absolute values involved, these quantities are nonsmooth functions of the state, which is undesirable 
for the purposes of numerical optimization.
Assuming that each vehicle respects the driving direction of the road, 
we have that $\cos \psi > 0$, so that we may omit the absolute value.
For $\sin \psi$, however, this is not the case, hence we will introduce 
an approximation.

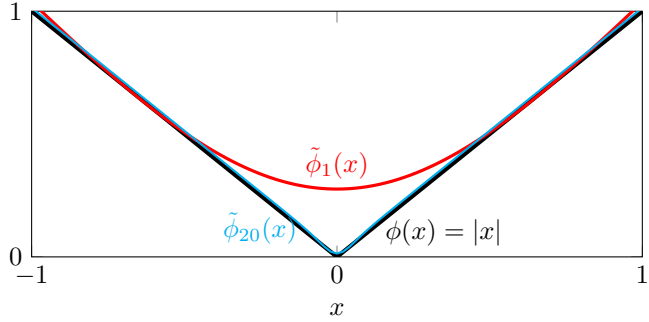
\begin{figure}[htb!]
    \centering
    \begin{tikzpicture}
        \def\lmbrt{0.5671}  
        \def\ftilde{\lmbrt/\k * (1 - tanh(\lmbrt)) + x * tanh(\k * x)}
        \begin{axis}[
                width=0.8\columnwidth,
                height=0.4\columnwidth,
                samples=101,
                xlabel={$x$},
                xtick={-1,0,1},
                ytick={0,1},
                xmin=-1, xmax=1.0, 
                ymin=0, ymax=1, 
                domain=-1:1,  
        ]
        \addplot[mark=none, ultra thick]    {abs(x)} node[midway, right, anchor=south west, xshift=0.5cm] {$\phi(x) = |x|$};
        \def\k{1}
        \addplot[mark=none, red, very thick]   {\ftilde}
        node[midway, right, anchor=south] {$\tilde{\phi}_{1}(x)$};
        \def\k{20}
        \addplot[mark=none, cyan, thick]   {\ftilde}
        node[midway, above, anchor=south east, xshift=-0.4cm] {$\tilde{\phi}_{20}(x)$};
        \end{axis}
    \end{tikzpicture}
    \caption{Smooth approximation \eqref{eq:smooth-abs} of $|x|$ for different values of 
    the smoothness parameter $k$.}
    \label{fig:abs-comparison}
\end{figure}

In particular, we will overapproximate $\phi: x \mapsto |x|$ by slightly modifying the result in \cite{bagul_SMOOTHTRANSCENDENTALAPPROXIMATION_2017}:
\begin{equation}   \label{eq:smooth-abs}
    \begin{aligned}
        &\tilde{\phi}_{k}(x) = b_k + x \tanh(k x), & b_k = \tfrac{\Wlamb(1)}{k} \left(1 - \tanh(\Wlamb(1))\right)
    \end{aligned}
\end{equation}
where $k$ is a parameter controlling the smoothness as illustrated 
in \cref{fig:abs-comparison} and $\Wlamb$ denotes the Lambert W-function \cite{corless_LambertWFunction_1996}. The higher $k$, the better the approximation, 
yet the larger the second derivatives around 0. In all experiments 
below, a value of $k=20$ is used.
Replacing now $|\sin \psi|$ with $\tilde{\phi}_{k}(\sin \psi)$ in expressions \eqref{eq:bounding-box}
for $\pmin$ and $\pmax$, these quantities become smooth functions of the augmented state $\xb$.
Note, finally, that $b_k$ is chosen  
such that $\tilde{\phi}_k(x) \geq |x|$ holds for all $x \in \Re$ (see \ref{app:smooth}), so that the full vehicle is always fully contained within the bounding rectangle, regardless of the approximation quality. 

Given two vehicles ($i \in \{0, 1\}$) with axis-aligned bounding boxes of lengths and widths $\id{l}{i}, \id{w}{i}$, it will be convenient to construct a single bounding box around the vehicle serving as the ``obstacle'' (by convention, vehicle $1$). For a position $\id{\pos}{1} = \trans{\smallmat{\id{\pos_{x}}{1} & \id{\pos_{y}}{1}}}$ of vehicle 1, this bounding box $\set{B}$ is defined as before, with
\[
    \pmin = \id{\pos}{1} - \tfrac{1}{2} \smallmat{\id{l}{0} + \id{l}{1} \\ \id{w}{0} + \id{w}{1}},
    \quad \pmax = \id{\pos}{1} + \tfrac{1}{2} \smallmat{\id{l}{0} + \id{l}{1} \\ \id{w}{0} + \id{w}{1}}.
\]
Collision avoidance constraints now amount to 
the statement $\id{p}{0} \notin  \set{B} $. This can be expressed in several ways. We discuss and explicitly compare two alternatives, which 
we refer to as the \emph{projection} and \emph{ellipse} formulation.

\paragraph{Projection formulation}

A direct formulation of the aforementioned constraints is simply to 
express the distance from the current position $\id{\pos}{0}$ to the 
expanded bounding box $\set{B}$, i.e., 
\begin{equation} \label{eq:rectangle-distance}
    \dist_{\set{B}}(\id{\pos}{0}) = \nrm{\id{\pos}{0} - \max\{ \pmin, \min\{\id{\pos}{0}, \pmax \} \}}\revision{}{_2}.
\end{equation}
Then the collision avoidance constraint will amount to 
$\dist_{\set{B}}(\id{\pos}{0}) > 0$. For numerical reasons, however, 
it is necessary to add some positive constant $\epsilon > 0$ and instead impose \eqref{eq:obstacle-avoidance-abstract} with 
\begin{equation} \label{eq:collision-distance}
    h(\xb) = \hproj(\xb) \dfn \epsilon - \dist_{\set{B}}(\id{\pos}{0})^2.
\end{equation}
Since this squared distance is only once differentiable, it may be
numerically beneficial to approximate this formulation by replacing the $\min$ and 
$\max$ operator in this constraint with a smooth function as 
follows. Note that $\max\{x,y\} = x \indicate_{x\geq y} + y (1-\indicate_{x\geq y})$, where 
\[ 
    \indicate_{x\geq y} = \begin{cases}
        1 & \text{ if } x \geq y \\
        0 & \text{ otherwise}
    \end{cases}
\]
is the indicator of the predicate $x \geq y$, which can be approximated using a sigmoid function, resulting in $\tilde{\indicate}_{x\geq y} = 
\sigma_{k}(x-y) = (1 + \exp (-k (x-y)))^{-1}$, with $k>0$ again some smoothness parameter (the larger, the better the approximation, but the worse the conditioning, due to increased curvature). We will refer to this approximated distance 
formulation as the \textbf{sigmoid approximation}. Experimentally (using 
the \IPOPT{} solver \cite{wachter_ImplementationInteriorpointFilter_2006}), 
this has been observed to improve the reliability with which
the online optimization problem is solved (see \cref{exp:experiment-1}).

\begin{figure}[ht!]
        \centering
        \begin{subfigure}[b]{0.49\columnwidth}
            \centering
            \includegraphics[width=\textwidth, trim=0 5cm 0 5cm, clip]{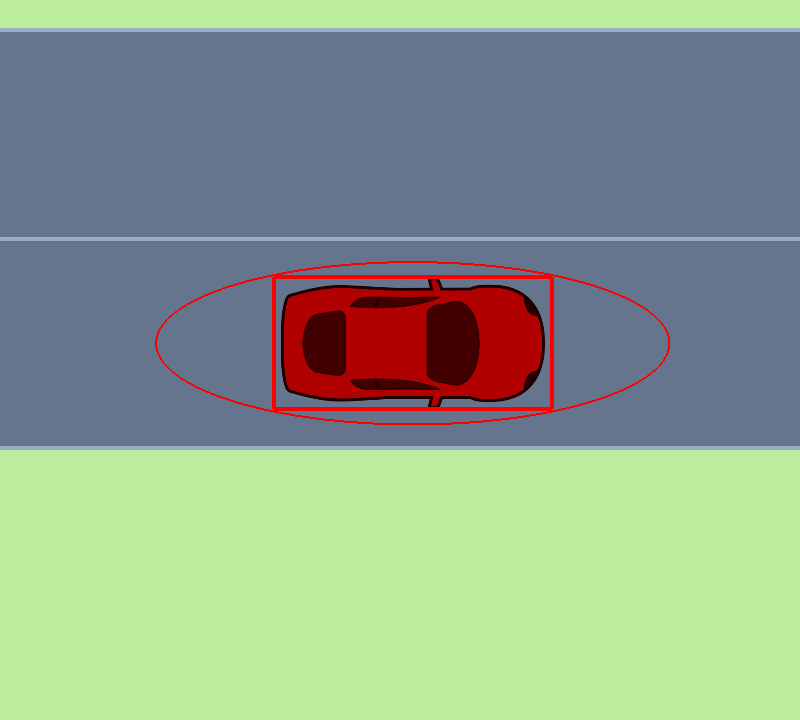}
            \caption{Elongated ellipse ($\gamma = 7$)}
        \end{subfigure} \hfill
        \begin{subfigure}[b]{0.49\columnwidth}
            \centering
            \includegraphics[width=\textwidth, trim= 0 5cm 0 5cm, clip]{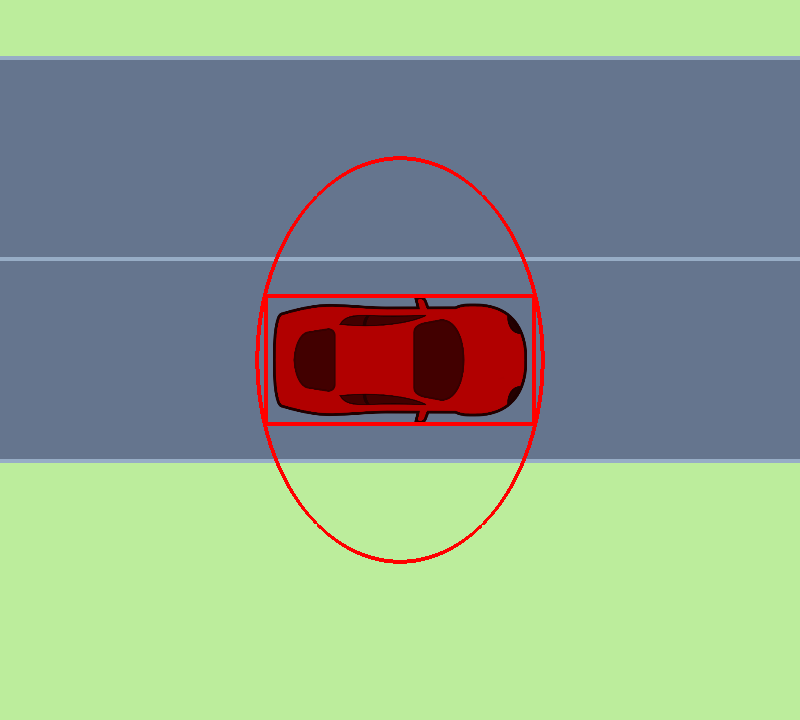}
            \caption{Widened ellipse ($\gamma = 0.5$)}
        \end{subfigure}
        \caption{Ellipsoidal sets around a given bounding box $\set{B}$, for different values of the elongation parameter $\gamma$.}
        \label{fig:ellipses}
    \end{figure}

\paragraph{Ellipse formulation} 
   As an approximation of the collision avoidance constraint, we fit an 
   ellipse 
   \[ 
       \set{E} = \left\{x \in \Re^2 \setsep \trans{\big(x-\id{p}{1} \big)} E \big(x- \id{p}{1}\big) \leq 1 \right\}
   \]
   around the bounding box $\set{B}$.
   Here, $E = \diag(\lambda_1, \lambda_2)$ is a positive definite diagonal matrix with eigenvalues 
   $\lambda_1$ and $\lambda_2 = \gamma \lambda_1$, where 
   $\lambda_1$ is computed such that the vertices of $\set{B}$ lie on the boundary of $\set{E}$, i.e., $\trans{\pmax} E \pmax = 1$, and $\gamma > 0$ is a freely chosen parameter controlling the elongation of the 
   ellipse (see \cref{fig:ellipses}). The resulting collision avoidance constraint then uses the function 
   \begin{equation} \label{eq:collision-ellipse}
   h(\xb) =  \hellipse(\xb) = 1 - \trans{(\id{p}{0} - \id{p}{1})}E{(\id{p}{0} - \id{p}{1})}.
   \end{equation}
    Note that since $\pmax$ and $\pmin$ are functions of the vehicle heading angle, $E$ is a function of the state $\xb$.
    \revision{However, with the simplified \emph{double integrator} dynamics, 
    the heading angle of the vehicle is fixed at 0, so that the dimensions
    of the bounding box $\set{B}$ are fixed and $E$ can be computed offline.
    Similarly, for modest heading angles, $E$ could be precomputed 
    for the bounding box that accounts for the largest allowed heading angle.
    }{However, for modest heading angles of the vehicle,
    $E$ could be precomputed for the bounding box that accounts for the largest allowed heading angle, 
    at the cost of only modest conservatism.
    }
    Nevertheless, since a closed-form expression for the parameters of $E$ exists, their computational cost is negligible, so we compute them online during our experiments, allowing us to account for the instantaneous
    orientation of the vehicles.
    In scenarios where tighter maneuvers at larger relative heading angles may be needed, a similar but less conservative formulation can be obtained by modeling a single target vehicle by several ellipses \cite{wang_NonGaussianChanceConstrainedTrajectory_2020}. 

\subsubsection{Arbitrarily oriented bounding boxes} If the bounding boxes 
of either of the vehicles are not mutually aligned, then the projection requires the following modifications. 

\begin{figure}[htb!]
    \centering
    \includegraphics{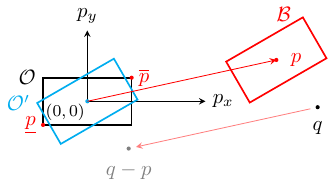}    
    \caption{Illustration of the distance to a generally oriented rectangle.}
    \label{fig:rectangles}
\end{figure}

\paragraph{Point-wise projection formulation}
We define $\set{O} = \lbrace v \mid \pmin \leq v \leq \pmax \rbrace$, where $\pmax = -\pmin = \nicefrac{1}{2}\trans{\smallmat{l & w}}$. Now the vehicle-aligned bounding box 
in a state $\trans{\smallmat{\pos & \psi & v}}$ is defined as
$\set{B} = \lbrace \pos + R(\psi) v \mid v \in \set{O} \rbrace$, where
\begin{equation*}
R(\psi) = \smallmat{
\cos(\psi) & -\sin(\psi)\\ \sin(\psi) & \cos(\psi)
},
\end{equation*}
as illustrated in \cref{fig:rectangles}. An expression for the 
distance is then obtained from \eqref{eq:rectangle-distance} as 
follows. 
If $\set{O}' = \{ R v \mid v \in \set{O} \}$, then for any point $q \in \Re^2$,
\begin{equation*}
    \dist_\set{B}^2(q) = \dist_{\set{O}'}^2(q-\pos). 
\end{equation*}
Due to the orthogonality of $R$:
\begin{equation*}
\begin{aligned}
    \dist_{\set{O}'}^2(q) &= \dist_\set{O}^2(\trans{R}q)\\
    \Rightarrow \dist_\set{B}^2(q) &= \dist_{\set{O}}^2(\trans{R}(q-\pos)).
\end{aligned}
\end{equation*}
Thus, the squared distance between a point $q$ and the rectangle $\set{B}$ is
\begin{equation*}
\begin{aligned}
    \dist_\set{B}^2(q) = \nrm{\trans{R}(q-\pos) - \min\lbrace  \max\lbrace \trans{R}(q-\pos), \pmin \rbrace , \pmax \rbrace}_2^2.
\end{aligned}
\end{equation*}
Similarly to \cite{evens_LearningMPCInteractionAware_2021}, we conclude that collision occurs whenever at least one corner of at least one vehicle is at distance zero from the other vehicle. For each target vehicle in the environment, this leads to 8 constraints similar to \eqref{eq:obstacle-avoidance-abstract} with functions $h$ of the form \eqref{eq:collision-distance}; one for every corner of every vehicle. The reduction of the approximation error thus comes at the cost of a moderate increase in computational demand of the constraint. 

It is worthwhile to note that many other collision constraint formulations exist \cite{hermans_PenaltyMethodBased_2018,  debrouwere_TimeoptimalPathFollowing_2013,zhang_optimization-based_2018}. In particular, preliminary experiments suggested that the method proposed in \cite{debrouwere_TimeoptimalPathFollowing_2013} performs 
similarly to the projection formulation. However, \revision{it introduces}{
    both \cite{debrouwere_TimeoptimalPathFollowing_2013} and 
    the well-known signed distance formulation \cite{zhang_optimization-based_2018} introduce} 
additional auxiliary variables for every node in the scenario tree, 
inducing a visible computational overhead. For this reason, \revision{it is}{they are} less suitable for planning over scenario trees and we have not included \revision{it}{them} in our final experimental comparison.

\subsection{Physical constraints}

We impose several constraints which enforce physically consistent or reasonable behavior and should therefore not be imposed 
probabilistically. The advantage of such constraints is that they 
can be imposed directly on the variables residing in the nodes of the 
scenario tree (see \eqref{eq:OCP} below), without requiring a risk measure and the corresponding 
risk reformulation described in \cite{sopasakis_RiskaverseRiskconstrainedOptimal_2019}. This 
reduces the number of auxiliary variables and constraints introduced
in the final optimal control problem.
We impose the following constraints in this manner. 

\paragraph{Road boundary constraints}
We impose that the lateral position of the ego vehicle remains between 
the road boundaries, i.e., constraints of the form 
$\id{\pos_y}{0} - \overline{c} \leq 0$ and $-\id{\pos_y}{0} + \underline{c} \leq 0$, 
where $\overline{c}, \underline{c}$ depend on the width of the road and 
the lateral dimensions of the axis-aligned bounding box.
\paragraph{Velocity constraints}
We enforce that all (longitudinal) velocities remain bounded from 
above by some 
positive limit $\vmax > 0$, and from below by 0.\revision{That is, $0 \leq \id{\vel}{0} \leq \vmax$ for the bicycle model and $0 \leq \id{\vel_x}{0} \leq \vmax$ for the double integrator model (assuming that $\vel_y$ is 
negligible, which is justified in the highway set-up under consideration).}{}    

For ease of notation, we concatenate these constraints and compactly write them as  
\[ 
    g_{\mathrm{phys}}(\xb) \leq 0, 
\] 
where $g_{\mathrm{phys}}: \Re^{\ns} \to \Re^{3}$ is an affine function.

\subsection{Control objective}

We conclude this section by specifying the control objective 
of the controller. The goal of the controller is to maintain a (given) constant velocity $\vref$, preferably in the rightmost lane. To this end, 
we define a desired state $\xref$ and specify the quadratic cost function $\ell: \Re^{\ns} \times \Re^{\na} \to \Re_+$ of the form
\[ 
    \ell(\xb,u) \dfn \trans{{(\id{x}{0} - \xref)}} Q (\id{x}{0} - \xref) + \trans{u} R u,
\]
where $Q \succeq 0$ and $R \succ 0$ (given in \Cref{tab:settings-ctrl}) are determined to
penalize in every time step 
\begin{inlinelist*}
\item the deviation of the lateral position of the host vehicle from the right lane center 
\item the longitudinal velocity of the host vehicle from some target value 
\item the total control effort.
\end{inlinelist*}
\revision{
If the double integrator model is used for the host vehicle, then 
$\xref = \trans{\smallmat{0 & 0 & \vref & 0}}$ is used, otherwise 
$\xref = \trans{\smallmat{0 & 0 & 0 & \vref}}$.
}{In other words, we set $\xref = \trans{\smallmat{0 & 0 & 0 & \vref}}$.}
The relative weighting was decided primarily to compensate for differences in scale of the different state/input variables, but could of course be tuned more specifically to reflect desired driving behavior. 
We choose the cost to be a function of the host vehicle state $\id{x}{0}$ 
only, but allow for more general costs as well, allowing to penalize
high relative velocities between vehicles in the environment, for instance, which may help reduce congestions. 

Finally, we remark that in general, the stage cost may also 
depend on the discrete mode $\md$, i.e., we have $\ell: \Re^{\ns} \times \Re^{\na} \times \W \to \Re_{+}$. This could be used to model changing preferences depending on the maneuver of surrounding vehicles. This poses no theoretical difficulties or additional computational cost. However, as this addition is not particularly meaningful for the case study at hand, we opt to 
simplify notation and disregard this dependence.
\section{Control methodology} \label{sec:methodology}

We will now formally describe the motion planning formulation.
As mentioned before, we formulate the decision-making 
problem as a distributionally robust optimal control problem 
\cite{schuurmans_GeneralFrameworkLearningBased_2021}, taking into 
account ambiguity in estimated transition probabilities. 
In the following, \revision{we compactly denote the joint (discrete-time) vehicle dynamics as $f: \Re^{\ns} \times \Re^{\na} \times \W \to \Re^{\ns}$}
{
    recall that $f: \Re^{\ns} \times \Re^{\na} \times \W \to \Re^{\ns}$ defined in \eqref{eq:dynamics-general} 
    denotes the joint (discrete-time) vehicle dynamics.
}

We will begin by introducing 
\revision{the related concepts \emph{risk} and \emph{ambiguity} in \cref{sec:risk-ambiguity}}{
the learning method in \Cref{sec:learning}. 
In \Cref{sec:ocp-scentree}, we describe how the learning task is embedded in the final 
control/decision-making problem that is solved online.
}

\revision{}{
\subsection{Learning ambiguity sets from data} \label{sec:learning}
\subsubsection{Independent data} \label{sec:iid}
We start by describing the learning problem in the simpler case of i.i.d. data. 
Let $(\W, \Fc, \prob)$ denote the probability space defined by 
the finite sample space $\W = \{ 1, \dots, \nModes\}$,
the set of all its subsets $\Fc$ and probability measure $\prob$.
A random variable $Z: \W \to \Re$ on this space can be identified with 
a vector $z = (z_i)_{i=1}^{\nModes} \in \Re^\nModes$, where $z_i \dfn Z(i)$.
Similarly, the distribution over $Z$ can be fully characterized by
a probability vector $p \in \simplex_\nModes$ 
with $p_i = \prob\{i\}$, $i \in \W$ the atomic probabilities, 
where $\simplex_\nModes \dfn \{ p \in \Re^{\nModes} \mid p \geq 0,  \trans{\1} p_i = 1 \}$ 
denotes the $\nModes$-dimensional probability simplex.
For a stochastic process $(\md_t)_{t=1}^{\hor}$, a similar notation is valid, 
taking the sample space to be $\W^\hor$.

For ease of exposition, 
suppose that $L: \Re^{n} \times \W \to \Re$ represents some random cost, and we are interested 
in minimizing its expected value 
(this will be extended to the multi-stage case in \cref{sec:ocp-scentree}):
\begin{equation} \label{eq:example-problem}
    \minimize_{u \in \Re^n} \E_{p}[ L\left( u, \md \right) ] 
    = \sum_{i \in \W} p_i L\left( u, i \right). 
\end{equation}
Of course, solving this problem requires knowledge of the probability distribution 
$p$, which we have assumed to be unavailable. 
If instead, we have access to a sample $\{ \hat{\md}_{k} \}_{t=1}^{t}$, 
drawn i.i.d. from $p$, we can compute the empirical estimate 
$\hat{p}_t \dfn t^{-1} \sum_{k=1}^{t} \e_{\md_k}$ and 
use this as an approximation for $p$. However, 
it is well-known that this approach tends to result in severe 
underestimations of the true expected costs, in particular for small sample sizes \cite{mohajerin_esfahani_data-driven_2018,parys_DataDecisionsDistributionally_2020}.
This phenomenon, which is akin to overfitting, can be 
mitigated in a systematic manner by additionally constructing 
a set of probability vectors which account for potential 
misestimations. Such a set is commonly referred to 
as an ambiguity set and can be constructed 
in a variety of manners, each valid for different settings 
and underlying assumptions
on the underlying system and data-generating distributions
\cite{yang_wasserstein_2018,mohajerin_esfahani_data-driven_2018,delage_DistributionallyRobustOptimization_2010,coppens_DatadrivenDistributionallyRobust_2021,schuurmans_learningBased_2020a}.

In this work, we build ambiguity sets by considering all probability distributions 
with some distance from the empirical distribution $\hat{p}_t$, 
expressed in the total variation (TV) metric. To do so, 
we leverage the following well-known concentration inequality.
\begin{lem}[{Total variation bounds \cite[Thm. A.6.6]{vaart_WeakConvergenceEmpirical_2000}}] \label{thm:concentration}
   Let $\{\md_k\}_{k=1}^{t}$ denote an i.i.d. sample drawn from a 
   distribution $p \in \simplex_{\nModes}$, and let 
   $\hat{p}_t \dfn t^{-1} \sum_{k=1}^{t} \e_{\md_k}$ denote the 
   empirical distribution. Then, for any confidence level $\conf \in (0,1)$,
    \begin{equation} \label{eq:radius}
        \prob[ \nrm{p - \hat{p}_t}_1 \leq r_{\conf, t}] \geq 1 - \conf, \text{ with } r_{\beta,t} = \sqrt{\tfrac{\nModes \log 2 - \log \conf}{t}}.
    \end{equation}
\end{lem}
This leads to the ambiguity set 
\begin{equation} \label{eq:ambiguity-set}
    \amb_t(\conf) = \left \{ \pi \in \simplex_{\nModes} \mid \nrm{\pi - \hat{p}_t}_1 \leq r_{\conf,t} \right \},
\end{equation}
which has the desirable property that the true distribution $p$ is captured by $\amb_{t}(\conf)$ with high probability, i.e.,
$\prob[ p \in \amb_t(\conf)] \geq 1 - \conf$.
Therefore, solving 
the distributionally robust counterpart of \eqref{eq:example-problem}
\begin{equation} \label{eq:DRO-simple}
    \hat{L} \dfn \min_{u \in \Re^n} \max_{p \in \amb_t(\conf)}\E_{p}[L\left( u , \md  \right) ], 
\end{equation}
to obtain a minimizer $u^\star_t$, ensures that with high probability, 
\begin{equation} \label{eq:certificate} 
    \E_{p}[L\left( u_t^\star, \md  \right) ] \leq \hat{L}.
\end{equation}
That is, the true, out-of-sample, expected cost will be no larger than the predicted cost.
Moreover, observe from \eqref{eq:radius} that the 
radius of the ambiguity set reduces at a rate of $\bigO(t^{-\nicefrac{1}{2}})$, which 
matches the expected convergence rate of the maximum likelihood estimator $\hat{p}_t$.

\begin{remark}[Recursive computation of \eqref{eq:ambiguity-set}] \label{rem:recursive-amb}
    We highlight that since $\amb_t(\beta)$ is fully described by its center $\hat{p}_t$ and its 
    radius $r_{\beta, t}$, it can be updated online at a negligible computational cost. 
    Indeed, given $\hat{p}_t$ and a new data point $\md_{t+1} \in \W$, it is well-known that $\hat{p}_{t+1}$ can be computed as
    \(
        \hat{p}_{t+1} = \frac{t \hat{p}_{t} + \e_{\md_{t+1}}}{t + 1},
    \)
    and $r_{\beta,t+1}$ is provided in closed form \eqref{eq:radius}.
\end{remark}

\subsubsection{Markovian data} \label{sec:markov}
Let us consider now the case where $(\md_t)_{t \in \N}$ is a Markov chain, 
defined on a probability space
$(\Omega, \Fc, \prob)$, which can be constructed as the product
space of the probability space introduced in \cref{sec:iid}.
We assume that the initial mode is known so that the Markov chain 
is fully specified by its transition probability matrix 
$\transmat = \left( \elem{P}{i}{j} \right)_{i,j \in \W} \in \Re^{\nModes \times \nModes}$ with 
$\elem{\transmat}{i}{j} \dfn \prob[ \md_{k} = j \mid \md_{k-1} = i ]$, $\forall k \in \N$.
Given a sample sequence $\{\hat{\md}_k \in \W \}_{k=1}^{t}$ from the Markov chain,
the sets $W_{t,i} \dfn \{ \hat{\md}_k \mid \hat{\md}_{k-1} = i, k \in \natseq{2}{t}\}$ for $i \in \W$
define i.i.d. data samples from the distribution $\row{\transmat}{i}$, i.e., the $i$'th row of
the transition matrix.
Using the procedure described in \cref{sec:iid}, we 
can for each mode $i \in \W$, construct an ambiguity set $\amb_{t,i}(\conf) \subseteq \simplex_{\nModes}$
defined as 
\begin{equation} \label{eq:mode-specific-ambiguity}
    \amb_{t,i}(\conf) = \{ p \in \simplex_{\nModes} \mid \nrm{ p - \row{\hat{\transmat}_t}{i} }_1 \leq r_{\conf,|W_{t,i}|}\left( \beta  \right)  \},
\end{equation}
with $\row{\hat{\transmat}_t}{i}$ the empirical distribution over the dataset $W_{t,i}$ and $r_{\conf,|W_{t,i}|}$
given by \eqref{eq:radius}, replacing $t$ with the cardinality of $W_{t,i}$.
Note that for all $t \in \N$ and $i \in \W$,
$\amb_{t,i}(\conf)$ is a random variable, since it is a function of 
the data sample $(\hat{\theta}_k)_{k = 1}^t$. 

\begin{remark}[Sample complexity] \label{rem:sample-complexity}
    It is possible to directly use 
    specialized concentration bounds derived for Markov chains (e.g., \cite{wolfer_MinimaxLearningErgodic_2019}), 
    instead of applying \cref{thm:concentration} for each individual row. 
    However, without further assumptions on the Markov chain (e.g., known mixing time), 
    these bounds provide no improvement over \cref{thm:concentration} and furthermore, 
    they would yield a single radius for all rows, whereas the proposed approach 
    allows the radii for the rows corresponding to more frequently active modes to decrease more rapidly.   
\end{remark}

}

\revision{}{
\subsection{Distributionally robust and risk-averse optimization} \label{sec:risk-measures}
Besides robustifying against estimation errors, the formulation \eqref{eq:DRO-simple} 
can also be interpreted as replacing the expectation in the cost function of the example problem \eqref{eq:example-problem} by 
a more general operator, called a \textit{risk measure}.
Given the finite sample space $\W$ with cardinality $|\W| = \nModes$, 
the outcomes of a random variable $Z: \W \to \Re$ can be represented by a vector $z \in \Re^{\nModes}$.
}
A risk measure $\rho: \Re^{\nModes} \to \Re$ then represents a mapping from an uncertain realization of $Z$ 
to an \emph{a priori} measure of its value. 
Trivial examples of risk measures are the expectation (corresponding to a \emph{risk neutral} attitude) 
and the $\max$ operation (corresponding to a \emph{robust} or fully \emph{risk-averse} attitude).
In particular, the class of \emph{coherent} risk measures are 
of interest for most problems involving decision-making \cite[\S 6.3]{shapiro_LecturesStochasticProgramming_2009}. 
These risk measures satisfy four additional properties, namely
\begin{inlinelist*}
    \item Monotonicity
     \item Convexity
     \item Translational invariance 
     \item Positive homogeneity, 
\end{inlinelist*} 
which ensure that they exhibit characteristics that one 
would intuitively expect, e.g., if a random variable $X$ is larger than $Y$ with probability 1, then the risk of $X$ must also be larger than the risk of $Y$;
\revision{}{For more details and intuition regarding these axiomatic properties, see, for instance \cite{majumdar_HowShouldRobot_2020,rachev2008advanced,chun_ConditionalValueatRiskAverage_2012}.
Furthermore, these properties endow coherent risk measures with sufficient structure to allow tractable reformulations (e.g., \cite{sopasakis_RiskaverseRiskconstrainedOptimal_2019}) 
of optimization problems involving them, despite their typical nonsmooth nature.
}

Coherent risk measures were originally used 
to represent the asymmetry in the preferences of decision makers.
In most cases, an incurred cost that is larger than the expected value carries more weight than one which is smaller by the same margin. 
This notion, which originated in finance and operations research, but 
has recently received increased interest in the control and robotics community \cite{dixit_RiskSensitiveMotionPlanning_2021,hans_RiskAverseModelPredictive_2020,carpin_RiskAversionFinite_2016,chapman_RiskSensitiveFiniteTimeReachability_2019}, 
considers the selection of the risk measure and its parameters as a degree of freedom for the designer.
Alternatively, however, coherent risk measures are intimately related to
\emph{distributionally robust} optimization through a 
well-known result, which we refer to as the \textit{dual risk representation} \cite[Thm. 6.4]{shapiro_LecturesStochasticProgramming_2009}.
This result states that one can associate with each coherent risk measure $\rho$, 
a nonempty, closed, and convex subset of the probability simplex (i.e., an ambiguity set)
$\amb_{\rho} \subseteq \simplex_\nModes$ such that 
\begin{equation*}
    \rho[z] = \max_{p \in \amb_{\rho}} \E_{p}[Z] = \max_{p \in \amb_{\rho}} \trans{p} z.
\end{equation*}
In other words, the risk measure $\rho$ can be written as a worst-case expectation 
over a set of probability distributions \revision{}{as in \eqref{eq:DRO-simple}}. 
We refer to $\amb_{\rho}$ as the ambiguity set \textit{induced by} the risk measure $\rho$.
\revision{}{
As a result of this equivalence, \eqref{eq:DRO-simple} inherits the 
favorable properties and intuitive interpretation of risk-averse optimization, 
but by virtue of \Cref{thm:concentration}, it has the 
added benefit that the risk measure is calibrated automatically based 
on the amount of data that is available, resulting in the out-of-sample 
certificate \eqref{eq:certificate}.
}
This in turn allows us to establish theoretical guarantees 
such as stability and constraint satisfaction of the closed-loop control system described in the next section
(see \cite{schuurmans_GeneralFrameworkLearningBased_2021} for details).

\revision{
 and then formulating the 
optimal control problem \eqref{eq:OCP} in \cref{sec:ocp-scentree}, which 
will be used in the case studies described further.

}{}

\subsection{Distributionally robust MPC over scenario trees} \label{sec:ocp-scentree}
\revision{
    We now move to the multi-stage setting, where we consider a 
    Markov chain $(\md_t)_{t \in \N}$, with $\md_t \in \W$, defined on a probability space $(\Omega, \Fc, \prob)$, which can be constructed as the product
    space of the probability space introduced in \cref{sec:risk-ambiguity}. 
}{
}

    \begin{figure}[ht]
       \centering
       \includegraphics[]{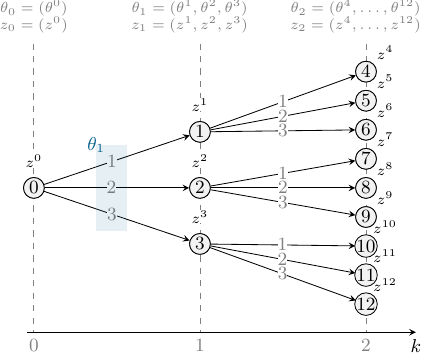}
       \caption{Scenario tree representation of some stochastic process $(z_t)_{t=0}^{2}$ adapted to the filtration generated by the Markov chain $(\md_t)_{t=0}^{2}$, with $\W = \{1,2,3\}$.}
       \label{fig:scentree}
    \end{figure}
\revision{}{
    \subsubsection{Scenario tree notation}
}
\revision{
    Let $\transmat = (\elem{\transmat}{i}{j})_{i,j \in \W}$ denote the transition probability matrix such that $\elem{\transmat}{i}{j} = \prob[\md_{t+1}=j \mid \md_t = i]$ for all $t \in \N$.
}{We now integrate the data-driven ambiguity sets obtained as described in \Cref{sec:learning} to
a multi-stage predictive control problem.}
    Since $\W$ is a finite set, the possible realizations of a \revision{}{mode} sequence
    $\seq{\md}{0}{\hor}$ can be enumerated and represented on a 
    \emph{scenario tree} \cite{pflug_MultistageStochasticOptimization_2014,schuurmans_GeneralFrameworkLearningBased_2021,sopasakis_RiskaverseRiskconstrainedOptimal_2019a}. 
    Similarly, 
    any stochastic process $(z_t)$ adapted to the filtration induced by $(\md_t)$ can be represented on the resulting scenario tree. We denote the value of $z_{t}$ corresponding to a
    node $\iota$ in the tree as $\nodevar{z}{\iota}$, as illustrated in \cref{fig:scentree}.
    The set of nodes in the tree are partitioned into time steps or \emph{stages}. The set of nodes at a stage $k$ is denoted by $\nodes{k}$,
    and similarly, for $k_0, k_1 \in \natseq{0}{\hor}$, with $k_1 > k_0$,  $\nodes{[k_0,k_1]} = \bigcup_{k=k_0}^{k_1} \nodes{k}$. Correspondingly, we have $z_t = (\nodevar{z}{\iota})_{\iota \in \nodes{t}}$.
    For a given node $\iota \in \nodes{t}$, $t \in \natseq{0}{\hor-1}$, we call a node $\iota_\pplus \in \nodes{t+1}$ that can be reached from $\iota$ in one step a \emph{child} node, denoted $\iota_{\pplus} \in \child{\iota}$. Conversely, we denote the (unique) 
    \emph{ancestor} node of a node
    $\iota \in \nodes{t}$, $t \in \natseq{1}{\hor}$ by $\anc{\iota} \in \nodes{t-1}$.
    \revision{}{We define an $n$-step \emph{ancestor} of a node $\iota$ recursively by $\anc[n]{\iota} \dfn \anc{\anc[n-1]{\iota}}$, with $\anc[0]{\iota} \dfn \iota$.}
    The nodes $\iota \in \nodes{\hor}$ have no child nodes and are called \emph{leaf nodes}. The unique node at stage $0$ has no ancestor and is called the \emph{root node}. 
    Finally, we will slightly abuse notation to write $\nodevar{z}{\child{\iota}} \dfn (\nodevar{z}{j})_{j \in \child{\iota}}$ for any non-leaf node $\iota$.
    \revision{}{
        \subsubsection{Multi-stage risk cost}
    }
    Using this construction, we can proceed to formulate the optimal 
    control problem for the ego vehicle. This problem, which will 
    take the form of a \emph{distributionally robust optimal control problem}, will be solved in receding horizon. At every time step, the realized value of $\md_t$ is observed, and the 
    involved ambiguity sets are updated accordingly, giving rise to a learning closed-loop control scheme.

    \revision{Let }{Consider some $N$-stage scenario tree with 
    $M = \sum_{k=0}^{N-1} | \nodes{k} |$ non-leaf nodes, and let}
    \[ 
        \mathbf{u}=\{\nodevar{u}{\iota} \in \Re^{\na} \mid \iota \in \nodes{[0, \hor-1]}\revision{}{\}}\revision{}{\in \Re^{M \na}}
    \] 
    denote a sequence of control actions over all non-leaf nodes the tree. 
    It is worthwhile to remark that the set of all such sequences suffice to represent all possible closed-loop $\hor$-step policies, without requiring a predetermined parametrization. 
    Consider the corresponding stochastic process $x_t \in \Re^{\ns}$, $t \in \N$, representing the system state and satisfying dynamics 
    \( 
        x_{t+1} = f(x_{t}, u_t, \md_{t+1}),
    \)
    or, in scenario tree notation, 
    \(
        \nodevar{x}{\iota} = f(\nodevar{x}{\anc{\iota}},
         \nodevar{u}{\anc{\iota}}, 
         \nodevar{\md}{\iota})
    \), for all $\iota \in \nodes{[1, \hor]}$. 
    Secondly, let $\ell: \Re^{\ns} \times \Re^{\na} \to \Re_+$ 
    and $\Vf: \Re^{\ns} \to \Re_+$
    denote some appropriate stage cost and terminal cost functions.
    \revision{}{
    To ease further notation, we define the mappings $\ell_t: \Re^{\na M} \to \Re_+^{|\nodes{t}|}$
    \begin{align} \label{eq:shorthands-costs}
            \ell_t(\bm u) &\dfn ( \ell\left( \nodevar{x}{\iota}, \nodevar{u}{\iota} \right) )_{\iota \in \nodes{t}} \text{ for } t = 0, \dots, \hor-1,\\
            \ell_N(\bm u) &\dfn ( \Vf\left( \nodevar{x}{\iota} \right) )_{\iota \in \nodes{\hor}}, 
    \end{align}
    where for any $\iota \in \nodes{t}$, $t \in \natseq{1}{N}$,
    $\nodevar{x}{\iota}$ depends on the (known) initial state $x_0$ and 
    the control actions
    $
    \{
        \nodevar{u}{\anc[k]{ \iota } } 
    \}_{k=1}^{t} 
    $. 
    }
    \revision{
    Given a sample sequence $\{\hat{\md}_k \in \W \}_{k=1}^{t}$ from the Markov chain,
    the sets $W_{t,i} \dfn \{ \hat{\md}_k \mid \hat{\md}_{k-1} = i, k \in \natseq{2}{t}\}$ for $i \in \W$ define i.i.d. data samples from
    the distribution $\row{\transmat}{i}$, i.e., the $i$'th row of
    the transition matrix.
    }{} Following the procedure described in \cref{sec:learning}, we 
    can for each mode $\md \in \W$, construct an ambiguity set $\amb_{t,\md}(\beta)$ containing possible distributions over its successor modes, 
    \revision{}{see \eqref{eq:mode-specific-ambiguity}}.
    \revision{}{Furthermore, since they can be precomputed at negligible computational cost (cf. \cref{rem:recursive-amb}),
    we can propagate future ambiguity sets (which are fully defined by their centers and radii) and store them on the scenario tree.
    Indeed, a node $\iota \in \nodes{t}$, $t \in [0, \hor-1]$, corresponds exactly to the event that the mode sequence $( \nodevar{\md}{\anc[k]{\iota}} )_{k=0}^{t}$
    is observed in the next $t$ time steps. Using these ``virtual'' data points, the center and the radius of the ambiguity set at node $\iota$ are uniquely determined.
    }
    \revision{yielding}{We denote the resulting ambiguity sets by} $\nodevar{\amb}{\iota}$ for all $\iota \in \nodes{[0, \hor-1]}$. 

\begin{figure}[htb!]
        \centering
        \includegraphics[]{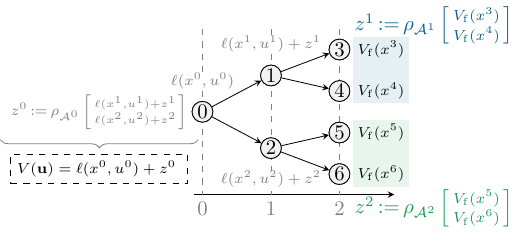}
        \caption{Graphical illustration of the computation of the cost
        in \cref{ex:example-cost} 
        on a scenario tree for a prediction horizon $\hor =2$, and $\W=\{1,2\}$.}
        \label{fig:cost_function_illustration}
    \end{figure}

    \revision{}{In doing so, we define at every time step, a \emph{conditional risk mapping}
    $\crm{t}[z_{t+1}]: \Re^{| \nodes{t+1}| } \to \Re^{|\nodes{t}|}$, defined as} 
    \cite{shapiro_LecturesStochasticProgramming_2009,sopasakis_RiskaverseRiskconstrainedOptimal_2019}
    \begin{equation} \label{eq:crm}
        \crm{t}[z_{t+1}] \dfn \big(\rho_{\nodevar{\amb}{\iota}}[\nodevar{z}{\child{\iota}}]\big)_{\iota \in \nodes{t}}.
    \end{equation}
    \revision{}{Intuitively, the conditional risk mapping $\crm{t}$ computes the risk the random outcome of a process at stage $t+1$, conditioned 
    on its value at stage $t$.}
    Using this construction, we define a multi-stage risk cost
    \revision{
        \begin{multline} 
        V(\mathbf{u}) \dfn
         \ell(x_0, u_0) + 
            \crm{0}\Big[
              \ell(x_1, u_1) + \crm{1} \big[ \dots \\
              \ell(x_{\hor-1}, u_{\hor-1}) + \crm{\hor-1}[\Vf(x_\hor)] \dots 
            \big ] 
            \Big].
        \end{multline}
    }{
    \begin{equation} \label{eq:total_cost}
        V(\mathbf{u}) \dfn
            \ell_0(\bm u) +
            \crm{0}\Big[
                \ell_1\left( \bm u \right)  + \crm{1} \big[ \dots 
                \ell_{\hor-1}(\bm u) + \crm{\hor-1}[\ell_\hor \left( \bm u \right) ]  \dots 
            \big ] 
        \Big],
    \end{equation}     
    with $\ell_t$, $t \in \natseq{0}{N}$ as in \eqref{eq:shorthands-costs}. This cost represents the worst-case expected cost over the 
    scenario tree of predicted $N$-step future scenarios.
    }
    \begin{example}[Cost function] \label{ex:example-cost}
        To illustrate the practical computation of the cost \eqref{eq:total_cost}, consider a prediction horizon of $\hor=2$ steps 
        and $\W = \{1,2\}$, resulting in the scenario tree depicted in 
        \cref{fig:cost_function_illustration}. 
        \revision{
            These values are stored in node 1 and 2, on which the risk mapping is conditioned. 
            These nodes additionally store the values of the stage cost $\ell(x_1, u_1)$ for the corresponding realizations, which are 
            subsequently added to the values $\nodevar{z}{i}$, $i=1,2$. 
            This process can be repeated, until the root node is reached, and 
            the value $V(\mathbf{u})$ is obtained. 
        }{In this case, the cost function \eqref{eq:total_cost} reduces to
        \begin{equation} \label{eq:total_cost_example}
            V(\mathbf{u}) =
                \ell_0(\bm u) +
                \crm{0}\Big[
                    \ell_1(\bm u) + \crm{1} 
                \big[ 
                        \ell_2(\bm u)
                \big] 
            \Big],
        \end{equation}     
        }
        The cost function can be evaluated through a backward substitution:
        \revision{}{
        The conditional risk mapping of the terminal cost 
        $\ell_2(\bm u) = \left( \Vf\left( \nodevar{x}{\iota} \right) \right)_{\iota = 3}^{6} \in \Re^4$
        is represented on this particular scenario tree as
        }
        \revision{$\crm{1}[\Vf(x_2)] = (\nodevar{z}{1}, \nodevar{z}{2})$}{
            \[ 
                \begin{aligned}
                  z_1 &\dfn \smallmat{\nodevar{z}{1} \\ \nodevar{z}{2}}
                    = \crm{1}[\ell_2 (\bm u)] \\ 
                        &= \smallmat{
                                \rho_{\nodevar{\amb}{1}}
                                \left[ 
                                     \Vf\left( \nodevar{x}{3}  \right), 
                                     \Vf\left( \nodevar{x}{4}  \right) 
                                \right]                                
                                \\
                                \rho_{\nodevar{\amb}{2}}
                                \left[ 
                                    \Vf\left( \nodevar{x}{5}  \right), 
                                     \Vf\left( \nodevar{x}{6}  \right) 
                                \right]
                            } & \text{(cf. $\eqref{eq:crm}$)}.
                \end{aligned}
            \]
        The values $\nodevar{z}{1}, \nodevar{z}{2}$ can then be assigned to the corresponding nodes 1, 2 on 
        the tree. Next, plugging these quantities into \eqref{eq:total_cost_example}, 
        we may define another auxiliary variable, 
            \[ 
                \begin{aligned}
                  \nodevar{z}{0} &\dfn \crm{0}
                  \left[ 
                    \ell_1(\bm u)  + z_1 
                  \right]
                  = \rho_{\nodevar{\amb}{0}}
                  \left[ 
                    \ell\left( \nodevar{x}{1}, \nodevar{u}{1} \right) + \nodevar{z}{1},  
                    \ell\left( \nodevar{x}{2}, \nodevar{u}{2} \right) + \nodevar{z}{2}  
                  \right],
                \end{aligned} 
            \]
        which is assigned to the root node 0, 
        so it can finally be combined with $\ell_0(\bm u)$ to  
        obtain $V(\bm u) = \ell(\nodevar{x}{0}, \nodevar{u}{0}) + \nodevar{z}{0}$, 
        as annotated on the bottom left of \cref{fig:cost_function_illustration}.
    }
    \end{example}
        
    \revision{}{
        \subsubsection{Ambiguous chance constraints}
    }
    Finally, following \cite{schuurmans_learningBased_2020a},
    collision avoidance constraints are imposed as \emph{ambiguous} conditional chance constraints.
    To this end, we utilize a popular coherent risk measure known as 
    the \emph{average value-at-risk}
    \cite{chapman_risk-sensitive_2019,sopasakis_RiskaverseRiskconstrainedOptimal_2019,hakobyan_risk-aware_2019}. 
    \revision{}{As before, let $Z$ denote a random variable, represented by its vector of realizations $z \in \Re^{\nModes}$, 
    and probability vector $p \in \simplex_{\nModes}$.}
    \revision{which, for}{For} a given risk level $\alpha > 0$
    \revision{and probability distribution $p \in \simplex_{\nModes}$}{},
    the average value-at-risk of $z$ is defined as
    \cite[\S 6.2.4]{shapiro_LecturesStochasticProgramming_2009}
    \begin{equation} \label{eq:avar-definition}
        \begin{aligned}
          \AVAR_{\alpha}^{p}[z] &= \min_{t \in \Re} t + \tfrac{1}{\alpha} \E_p \left[[Z-t]_+ \right] \\ 
                                &= \min_{t \in \Re} t + \tfrac{1}{\alpha} \langle p, [z-t]_+ \rangle.
        \end{aligned}
    \end{equation}
    \revision{
    Constraining this function to be below $\alpha$ is equivalent to asserting that $\AVAR_{\alpha}^{p}[Z] \leq 0$ \cite[\S 6.2.4]{shapiro_LecturesStochasticProgramming_2009}.
    }{It can be shown that, $\AVAR_{\alpha}^{p}[z] \leq 0 \implies \prob[Z > 0] \leq \alpha$\footnote{See \ref{app:background-chanceconstraints} for additional details.},
    and furthermore, the former inequality can be exactly reformulated as a smooth constraint by introducing some additional auxiliary variables \cite{sopasakis_RiskaverseRiskconstrainedOptimal_2019}.}
    Besides resulting in a safe, continuous surrogate of the original chance constraint,
    the average value-at-risk has the additional practical benefit of penalizing larger constraint violations and thus providing more informative gradients to local optimization solvers. 
    
    The evaluation of \eqref{eq:avar-definition}, however, requires knowledge of $p$,
    but our controller only has access to the estimated ambiguity sets \eqref{eq:mode-specific-ambiguity}.
    For this reason, we impose
    constraints of this type robustly with respect to misestimation of 
    $p$. In particular, this leads to the collision constraints
    \begin{equation} \label{eq:collision-constraint}
        \revision{
        \nodevar{\bar \rho}{\iota}\left[h \big(\nodevar{x}{\child{\iota}} \big)\right] \dfn \max_{p \in \nodevar{\amb}{\iota}} \AVAR_{\alpha}^{p}\left[h \big(\nodevar{x}{\child{\iota}} \big) \right] \leq 0,
        }{
        \nodevar{\bar \rho}{\iota}\left[
            \Big( h \big(\nodevar{x}{i} \big) \Big)_{i \in \child{\iota}}
        \right] 
        \dfn \max_{p \in \nodevar{\amb}{\iota}} \AVAR_{\alpha}^{p}\left[
            \Big( h \big(\nodevar{x}{i} \big) \Big)_{i \in \child{\iota}} \right] \leq 0,
        }
    \end{equation}
    for all $\iota \in \nodes{[0,\hor-1]}$, 
    where $h: \Re^n \to \Re$ is some function of the state that represents a negative distance. \Cref{sec:collision-avoidance-forms}
    describes several possible choices for the function $h$.
    
    \revision{}{\subsubsection{Full optimal control problem}}

    For a given state $x$, we can now combine these components together with the previously introduced constraints in \Cref{sec:modelling} into the optimal control problem
    \begin{equation} \label{eq:OCP}
        \begin{aligned}
        & \minimize_{\mathbf{u}}&& \cost(\mathbf{u})\\
        &\stt&&\nodevar{x}{0} = x, \\ 
        &&& \nodevar{x}{\iota} = f(\nodevar{x}{\anc{\iota}},
        \nodevar{u}{\anc{\iota}}, 
        \nodevar{\md}{\iota}), \; \forall \iota \in \nodes{[1, \hor]}, \\
        &&& \nodevar{\bar \rho}{\iota}\left[
            \Big( h \big(\nodevar{x}{i} \big) \Big)_{i \in \child{\iota}}
        \right] \leq 0, \; \forall \iota \in \nodes{[0, \hor-1]},\\
        &&& \underline{u} \leq u \leq \bar{u}, \; \forall \iota \in \nodes{[0,\hor-1]},\\
        &&& g_{\textrm{phys}}(\nodevar{x}{\iota}) \leq 0, \; \forall \iota \in \nodes{[0, \hor]}.
        \end{aligned}
    \end{equation}
    \revision{}{
    \begin{remark}[Feasibility]
        In order to guarantee that a control action is computed at all times, it is common 
        in practice to relax the constraints by introducing slack variables (see e.g., \cite{zhang_optimization-based_2018}), resulting in a soft-constrained 
        version of \cref{eq:OCP}. 
        An alternative option is to hand-craft a back-up controller,
        which gets invoked whenever the optimal control problem is infeasible 
        \cite{brudigam_SafeStochasticModel_2022,nair_CollisionAvoidanceDynamic_2022}.
        However, since the aim of this work is to study the behavior of \eqref{eq:OCP} in
        closed-loop, we consider the design of additional safeguards to be beyond the scope of this work.
    \end{remark}
    }

    In typical \ac{MPC} fashion, our proposed controller involves repeated application of the following steps:
    \begin{inlinelist*}
        \item solve \eqref{eq:OCP} for the current state $x$
        \item apply the resulting control action in the root node
        \item measure/observe the new state and mode
        \item update the ambiguity set parameters and repopulate the scenario tree.
    \end{inlinelist*}
    
    The ambiguity set \eqref{eq:ambiguity-set} belongs to the 
    general set of \emph{conic-representable} risk measures, which 
    allows us to tractably reformulate the \ac{OCP} \eqref{eq:OCP} 
    as a standard nonlinear program,
    using duality-based techniques described in \cite{sopasakis_RiskaverseRiskconstrainedOptimal_2019,schuurmans_GeneralFrameworkLearningBased_2021,coppens_DatadrivenDistributionallyRobust_2021}.
\section{Numerical case studies} \label{sec:case-study}
\subsection{Exploratory experiments} \label{sec:exploratory}

\begin{figure*}[ht!]
    \centering
    \includegraphics[width=\textwidth]{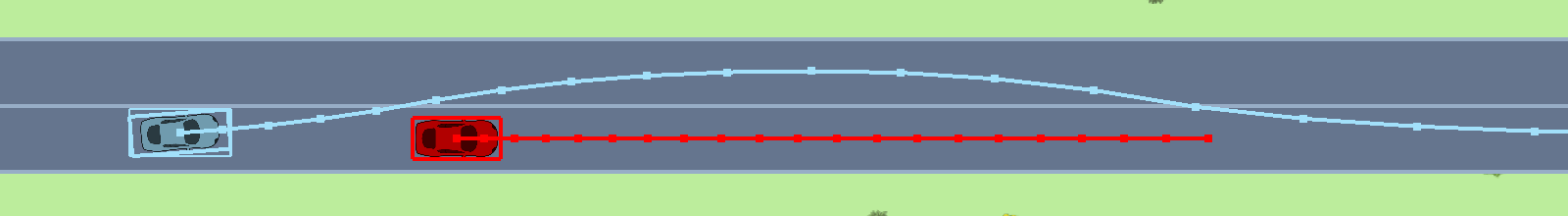}
    \caption{Scene layout for the experiments in \Cref{sec:exploratory,sec:experiment-topology}. (Blue: ego vehicle; red: target vehicle)}
    \label{fig:overtaking-layout}
\end{figure*}

This section provides some exploratory and qualitative simulation
results which motivate some of the decisions made for the 
configurations of further experiments. Unless stated otherwise, 
the sampling time used for these experiments is $\Ts = \SI{0.2}{\second}$.
The reference velocity was kept at $\vref = \SI{30}{\meter\per\second}$. 
For other settings, we refer to \ref{app:settings}.

All nonlinear programs are solved using \IPOPT{} \cite{wachter_ImplementationInteriorpointFilter_2006}, using the \CASADI{} \cite{Andersson2018}
interface in Python.  

\begin{experiment}[Collision avoidance formulations] \label{exp:experiment-1}

As a first comparison of the effectiveness of the above collision-avoidance constraints, 
we run a closed-loop simulation 
of an overtaking scenario with a single target vehicle, shown 
in \cref{fig:overtaking-layout}. For the time being, we employ
\textbf{deterministic} model predictive control, i.e., we keep the mode of the (single) target vehicle fixed to the right lane, both for the predictive model and the simulation model.
This allows us to isolate the effects of the choices regarding collision avoidance formulations. 

\Cref{fig:trajectories-collision-constraints} gives an overview of the 
resulting trajectories of the ego vehicle with the different formulations. 
For the projection formulation, the solver occasionally returned 
an infeasible solution. In this experiment, we accepted the solution
regardless, allowing the controller to correct using feedback in the 
next time instances.

\revision{
\begin{table}[ht!]
    \centering
    \caption{Infeasible instances per configuration (out of 50).}
    \label{tab:infeasibility-exp1-old}
     \begin{tabular}{@{}llllll@{}}
    \toprule
    & &  \multicolumn{4}{@{}c@{}}{Projection} \\
    \cmidrule{3-6}
    & Ellipse & \parbox{1cm}{AA$^{\hyperlink{AA}{\star}}$\\(exact)} & \parbox{1cm}{AA$^{\hyperlink{AA}{\star}}$\\(smooth)} &  exact &  smooth\\
    \midrule
    Double integrator &0 & 3& 0& -& -\\
    Bicycle     &0 & 19 & 0 & 1 & 1\\  
    \bottomrule
    {\footnotesize \hypertarget{AA}{$^\star$\textit{Axis-Aligned}}}
    \end{tabular}
 \end{table}
}{
\begin{table}[ht!]
    \centering
    \caption{Infeasible instances per configuration (out of 50).}
    \label{tab:infeasibility-exp1}
    \begin{tabular}{@{}llllll@{}}
    \toprule
    &\multicolumn{4}{@{}c@{}}{Projection} \\
    \cmidrule{3-6}
    \revision{}{Formulation} & Ellipse & \parbox{1cm}{AA$^{\hyperlink{AA}{\star}}$\\(exact)} & \parbox{1cm}{AA$^{\hyperlink{AA}{\star}}$\\(smooth)} &  exact &  smooth\\
    \midrule
    \revision{}{Failure cases} & 0 & 19 & 0 & 1 & 1\\  
    \bottomrule
    {\footnotesize \hypertarget{AA}{$^\star$\textit{Axis-Aligned}}}
    \end{tabular}
\end{table}
}
 
In \Cref{tab:infeasibility-exp1}, the number of infeasible instances are found for the different collision avoidance constraints.\revision{ (columns) and ego vehicle dynamics (rows)}{}
This leads to several conclusions:
\begin{inlinelist}
   \item The ellipsoidal and smoothened, axis-aligned projection-type collision avoidance constraints perform the most reliably, yielding no infeasible solutions. We conclude that this formulation is most numerically well-behaved, as infeasibilities are not caused by model mismatch or other sources of uncertainty;
   \item The sigmoid approximation for the projection-type constraints 
   decreases the number of infeasible instances significantly;
   \item The exact projection-type constraint leads to infeasible solutions at a rate that prohibits corrections using feedback, leading 
   to a collision in the closed-loop trajectories; 
   \item From \cref{fig:trajectories-collision-constraints}, the conservatism introduced by either the ellipsoidal constraints 
   or the smoothing approximation seems insignificant. 
\end{inlinelist}

\begin{figure}[htb!]
    \centering
    \revision{
    \begin{subfigure}{\columnwidth}
        \includegraphics{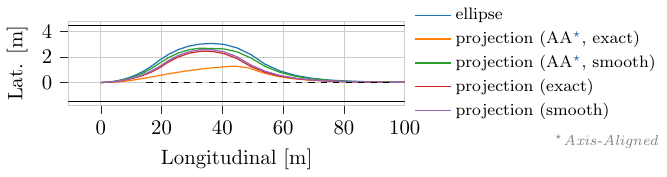}
        \caption{Bicycle model.}
    \end{subfigure}
    }{
    \includegraphics{qualitative_nonlin--collision_positions.pdf}
    } 
    \caption{Overtaking trajectories for different collision-avoidance 
   formulations. The colored lines indicate the ego vehicle trajectories for different collision avoidance constraints. The dashed line indicates the target vehicle positions. (Best viewed in color.)}
   \label{fig:trajectories-collision-constraints}
\end{figure}

This experiment, combined with similar test runs yielding observations consistent with the above conclusions motivates the use of the \textbf{ellipsoidal
constraints} by default. 

\end{experiment}

\begin{experiment}[Ellipsoidal parameter tuning]
    A remaining degree of freedom of the ellipsoidal constraint is 
    the choice of the tuning parameter $\gamma$, which determines the 
    ``elongation'' of the ellipsoidal shape. The 
    lower this value, the wider the ellipse. If it is chosen too low,
    an overtaking maneuver will be rendered impossible, as illustrated 
    in \cref{fig:gamma1}. On the other hand, if an excessively large 
    value is chosen, the time spent in the non-preferred lane during 
    an overtaking maneuver increases, with an increased cost as a result. 
    This trade-off is illustrated in \cref{fig:cost-vs-elongation}. 
    For the current cost function, values between 7 and 15 were observed 
    to provide the best performance.
   
    \begin{figure}[ht!]
       \centering
       \includegraphics{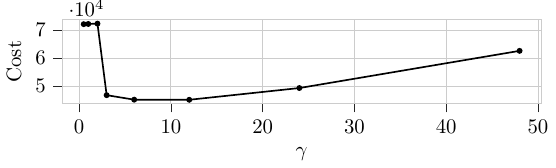}
       \caption{Closed-loop cost versus the elongation parameter $\gamma$.}
       \label{fig:cost-vs-elongation}
    \end{figure}

    \begin{figure}[ht!]
        \setlength{\trimh}{6cm}
        \begin{subfigure}{0.48\textwidth}
            \begin{minipage}{\textwidth}
               \includegraphics[width=\textwidth,trim=0 {\trimh+5cm} 0 {\trimh}, clip]{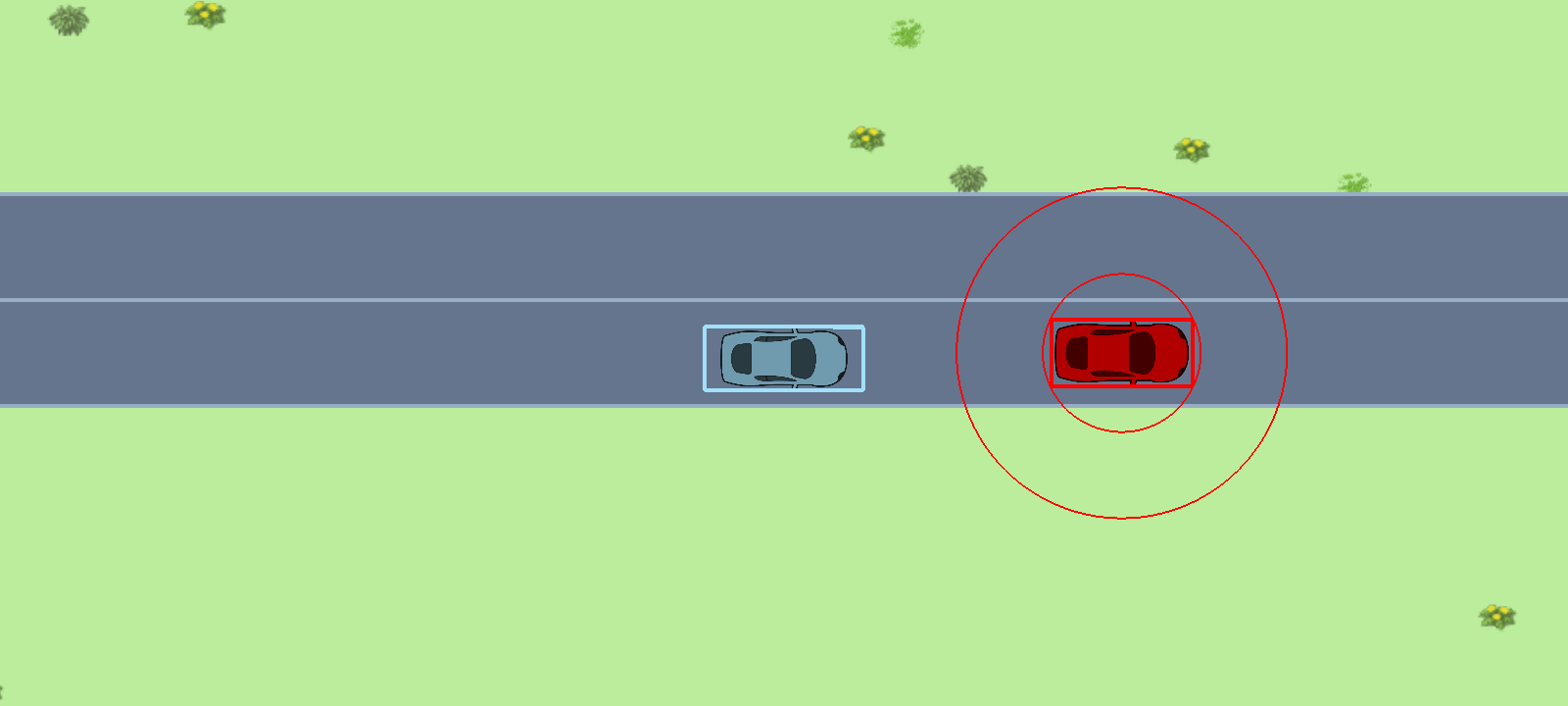}
  
               \includegraphics[width=\textwidth, trim=0 {9cm} 0 {\trimh+2cm}, clip]{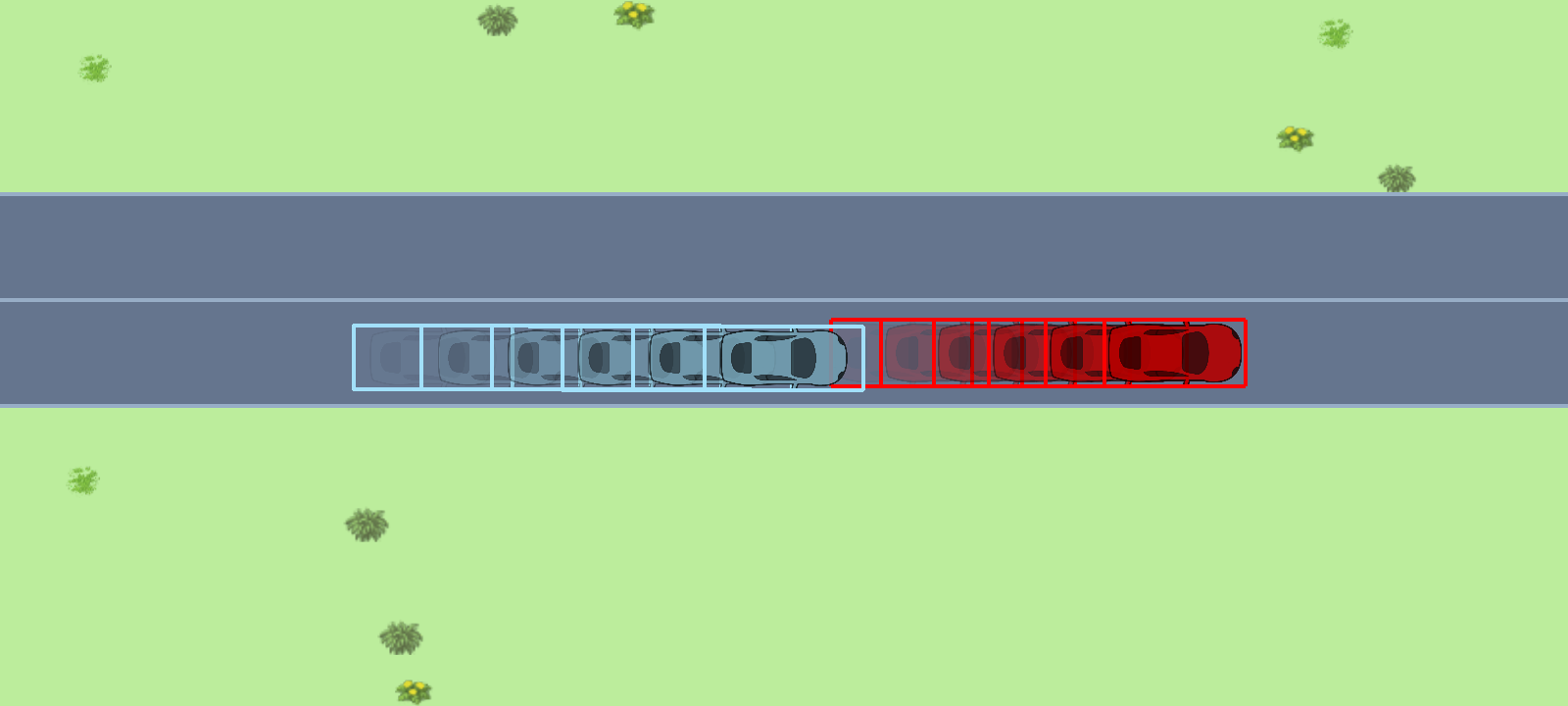}
            \end{minipage}
            \caption{$\gamma=1$}
            \label{fig:gamma1}
        \end{subfigure}
       \begin{subfigure}{0.48\textwidth}
          \begin{minipage}{\textwidth}
             \includegraphics[width=\textwidth, trim=0 {9cm} 0 {\trimh}, clip]{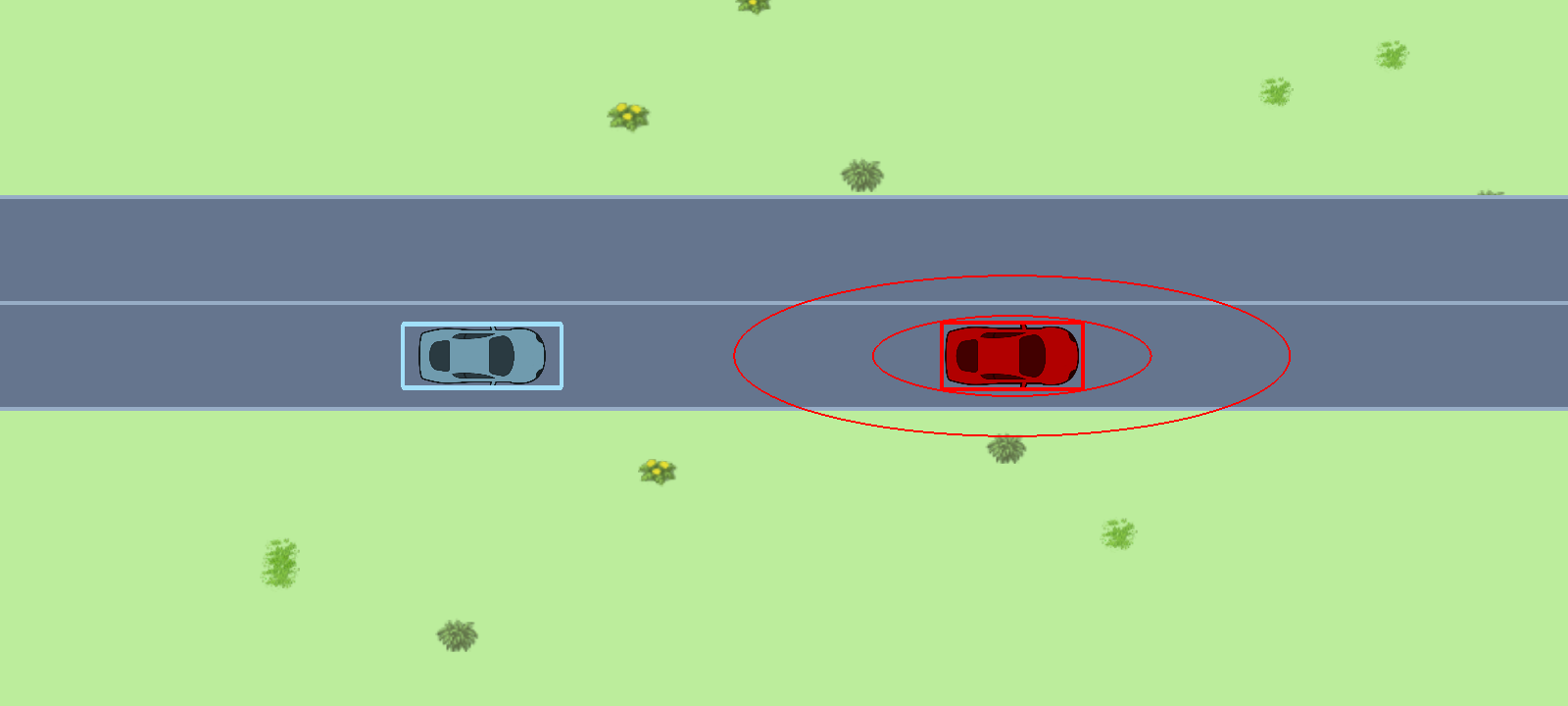}

             \includegraphics[width=\textwidth,trim=0 {9cm} 0 {\trimh}, clip]{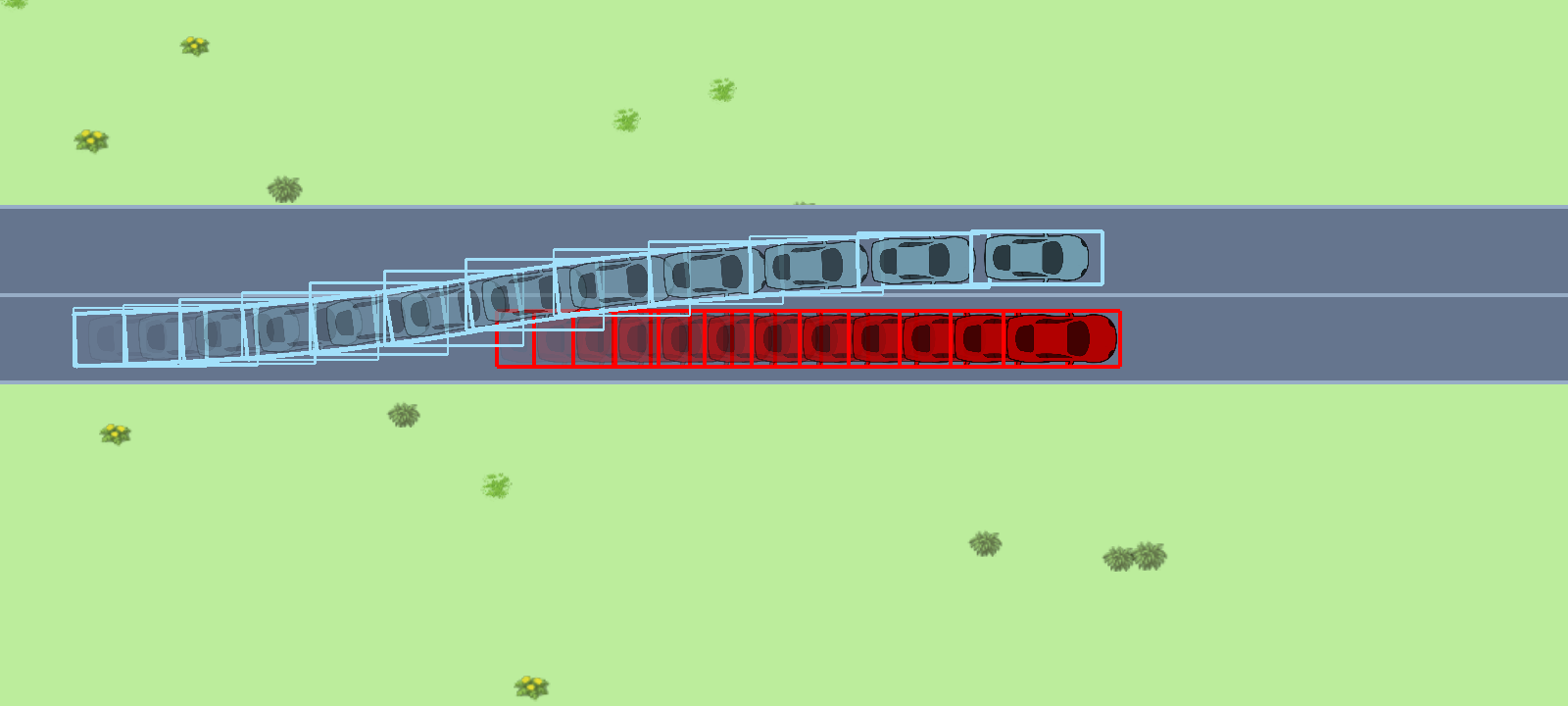}
          \end{minipage}
          \caption{$\gamma=12$}
          \label{fig:gamma12}
       \end{subfigure}
       \caption{Illustration of the ellipsoidal constraint sets and closed-loop trajectories for different values of the 
       elongation parameter $\gamma$. The outer ellipse corresponds to the extended rectangle and delineates the constraint set for the 
       center point of the ego vehicle. (Blue: ego vehicle; red: target vehicle.)}
    \end{figure}
\end{experiment}

\subsection{Effects of sample-size and scenario tree topology on conservatism} \label{sec:experiment-topology}
We now move on to the setting where a (nontrivial) scenario tree is built for the predictive model and the full learning distributionally robust \ac{MPC} framework
is used for control. The following experiments illustrate the effects of two important parameters that control risk-aversion in this setting.

\subsubsection{Deterministic target vehicle}
We first study the behavior of the controller in a 
situation where the true behavior of the target vehicle is 
rather safe and predictable, i.e., it regulates to a fixed 
reference velocity in the right lane and performs no lane-change 
maneuvers, i.e., the transition kernel governing the target 
vehicle driving style
in this experiment is $\transmat = \smallmat{1 & 0\\ 1&0}$.
Recall that the modes in this set-up represent a target lane, i.e., element $\elem{P}{i}{j}$ in $\transmat$ represents the probability that 
at any time, the target vehicle will change its desired lane to lane $j$ given that it is currently driving towards lane $i$. Thus, with this instance of $\transmat$, the target vehicle will drive towards the 
center of the right lane with probability 1, regardless of the initialization of the Markov chain. 
The overall setup of the scenario is the same as in \Cref{exp:experiment-1}.
This scenario is representative of a large fraction of expected driving situations and is therefore an interesting case to consider. 

Under these conditions, it is evident that a deterministic MPC scheme would 
be optimal, since the mode sequence is fixed a priori and therefore,
the deterministic predictions are exact. In these experiments, we will therefore refer to the deterministic controller as \textit{prescient}.
Since by construction of the driving scenario, the behavior of the target vehicle is predictable and safe for an overtaking maneuver, the interesting question
is: 
\emph{``How conservative is the learning distributionally robust MPC scheme in an inherently safe scenario, and does this lead to impractical behavior?''}

\begin{experiment}[Conservatism versus branching horizon] \label{exp:horizon-1}
    The current experiment investigates the effect of the following two 
    quantities on the conservatism of the collision avoidance constraints \eqref{eq:collision-constraint}:
    \begin{inlinelist}
       \item the amount of observed data from the target vehicle; and 
       \item the branching horizon used in the scenario tree.
    \end{inlinelist}
    
    We simulate the \ac{MPC} scheme in closed loop for 
    the overtaking scenario and record its achieved cost by 
    evaluating the stage cost at every time step and summing over 
    the simulation run.
    To simulate the behavior of the controller after observing 
    the target vehicle for some time, we sample a sequence 
    of length $\nsample$ from the governing Markov chain to 
    train the learning controller. Note that since we initialize the empirical distribution to the uniform distribution and the experiment is constructed such that $\md_t = 1, \forall t \in \N$, the estimated probability matrix will be identically
    $\hat{\transmat} = \smallmat{1 & 0 \\ 0.5 & 0.5}$, for any sample size 
    larger than 1. Therefore, any difference in the behavior stems from 
    the confidence on these estimates, as measured by the radii
    of the ambiguity sets. 

    \cref{fig:cost-vs-horizon} shows estimated closed-loop costs over 
    a simulation versus this sample size $\nsample$.
    As the sample size grows and the radii of the ambiguity sets shrink
    according to \eqref{eq:radius},
    the controllers become more confident about the prediction that a sudden 
    lane change will not occur and decide to perform an overtaking maneuver, which results in the sudden decrease in costs in the figure.
    Note that the smaller the branching horizon, the more quickly this
    threshold is reached, as fewer potentially dangerous future situations are taken into account during prediction. 
    
    However, after roughly 50 data points -- corresponding to $10$ seconds of observation -- even the most conservative design 
    is able to perform the overtaking maneuver without violating its constraints, achieving the same closed-loop cost as the \emph{prescient} controller. Meanwhile, this controller does plan with respect to a considerably more detailed model, which robustifies it against use cases where a sudden lane chan\revision{}{g}e would spuriously occur. In contrast, a robust formulation,
    where the collision avoidance constraints are imposed in the worst-case realization, an overtaking maneuver is not feasible, as illustrated by the behavior of the \ac{DR} controller at sample size $\nsample=0$. 

    \begin{figure}[ht!]
        \begin{subfigure}[b]{0.6\columnwidth}
            \centering
            \includegraphics[]{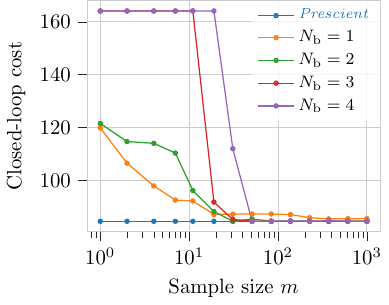}
            \caption{Cost versus sample size.}
            \label{fig:cost-vs-data-deterministic}     
        \end{subfigure}
        \begin{subfigure}[b]{0.35\columnwidth}
            \centering
\newlength{\vspacingnodes}
\setlength{\vspacingnodes}{1pt}

\newlength{\hspacingnodes}
\setlength{\hspacingnodes}{5pt}
{\small

\includegraphics[]{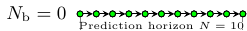}
\vspace{2\vspacingnodes}
\includegraphics[]{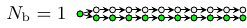}
\vspace{2\vspacingnodes}
\includegraphics[]{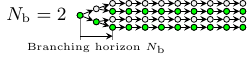}
\vspace{2\vspacingnodes}
\includegraphics[]{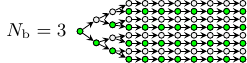}
\vspace{2\vspacingnodes}
\includegraphics[]{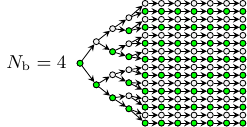}
}
            \caption{Scenario tree topologies.}
            \label{fig:topologies1}     
        \end{subfigure}
        \caption{Closed-loop costs versus sample size and scenario tree topologies corresponding to different branching horizons $\bhor$
        and fixed prediction horizon $\hor = 10$ 
        in the deterministic setting of \Cref{exp:horizon-1}. The colored nodes indicate modes where 
        the target vehicle is predicted to move towards the right lane.}
        \label{fig:cost-vs-horizon}
    \end{figure}
\end{experiment}

\begin{experiment}[Prediction breadth versus depth] \label{exp:horizon-2}
    In \Cref{exp:horizon-1}, an increasing branching horizon was demonstrated to rather quickly yield optimal behavior in the case where 
    the true underlying system is not exhibiting low-probability 
    switching behavior, even though it improves resilience of the 
    control scheme to unexpected behaviors.

    Naturally, however, an increased branching horizon also induces larger scenario trees, and therefore higher computational costs.  
    For this reason, we repeat \Cref{exp:horizon-1} with 
    (approximately) constant scenario tree sizes. That is, the prediction horizon (\emph{``depth''} of the prediction) is traded off with the branching horizon (\emph{``breadth''} of the prediction) such that the 
    total size of the scenario tree does not change significantly --- see \cref{fig:topologies2} for an illustration of the resulting scenario trees. In doing so, we investigate whether it is beneficial to 
    increase the detail of the prediction at a given stage at the 
    cost of a reduced prediction horizon, given a fixed computational budget.  

    \cref{fig:cost-vs-horizon-fixedsize} shows that in this 
    setup a degradation in the performance can indeed be observed 
    whenever the trade-off is made too heavily in favor of \textit{breadth} rather than \textit{depth}. 
    Although at sufficiently large sample sizes, the controllers 
    with short prediction horizons are still able to perform an overtaking maneuver (note the sudden decrease in closed loop costs), they do so at a significantly slower pace, as only a part of the maneuver can be predicted at each time instance. This naturally leads to higher costs.
    Note, however, that until a branching horizon of 2, there does not yet 
    seem to be a significant degradation in performance as a result of 
    the shortened horizon.
    
    This suggests that there exists a minimal prediction horizon that 
    is required to allow the prediction of the optimal maneuver. As long 
    as the prediction horizon is longer than this value, the optimal behavior is 
    obtained (given enough time to learn the governing distributions). 
    \revision{
    By consequence, it is expected that as a larger sampling time is used, 
    and thus the elapsed \textit{real} time per discrete time step is increased, 
    the lower this required horizon. 
    }{
    Since for a fixed number of time steps, the length of the preview window (expressed in true seconds) 
    is proportional to the sampling time, it is expected that the
    required prediction horizon to predict the full overtaking decreases as $\Ts$ increases.
    }
    Indeed, by increasing the sampling horizon to $\Ts=\SI{0.3}{\second}$, as presented in \cref{fig:cost-vs-horizons-fixed-ts-03}, the discrepancy between the results for different values of $\bhor$ is 
    reduced considerably. It is however important to emphasize that despite the observed trend, the closed-loop cost need not increase monotonically with the branching horizon, as one may suspect from \cref{fig:cost-vs-horizon-fixedsize} alone. Indeed, in the case of \cref{fig:cost-vs-horizons-fixed-ts-03} for instance, the controller using $\bhor = 2$ outperforms the one using $\bhor =1$ at smaller sample sizes. It is therefore advisable in practice to fine-tune the exact choice of $\bhor$ to the application at hand. 

    We conclude that depending on the timescale of the use case, a 
    minimal prediction horizon may be required in order to 
    achieve satisfactory performance. In such cases, it may be beneficial
    to reduce the branching horizon by a number of steps to make this possible. An important caveat here, is that if a terminal constraint is
    imposed, it needs to be imposed at stage $\bhor$ for it to 
    guarantee recursive feasibility of the scheme. As a result, $\bhor$ should be as large as possible in order to maximally enlarge the region of attraction.

    \begin{figure}[ht!]
        \centering
        \begin{subfigure}[b]{0.45\columnwidth}
            \includegraphics{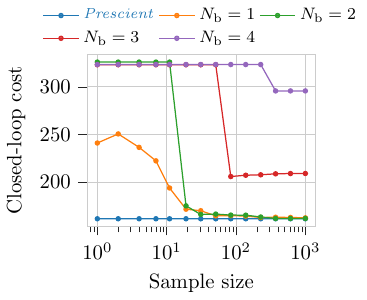}
            \caption{Cost versus sample size}
            \label{fig:cost-vs-horizon-fixedsize-1}
        \end{subfigure}
        \begin{subfigure}[b]{0.5\columnwidth}
\ifdefined\vspacingnodes\else\newlength{\vspacingnodes}\fi 
\setlength{\vspacingnodes}{1pt}

\ifdefined\hspacingnodes\else\newlength{\hspacingnodes}\fi 
\setlength{\hspacingnodes}{2pt}
{\small

\includegraphics[]{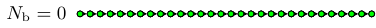}
\vspace{2\vspacingnodes}
\includegraphics[]{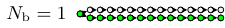}
\vspace{2\vspacingnodes}
\includegraphics[]{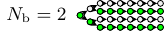}
\vspace{2\vspacingnodes}
\includegraphics[]{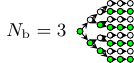}
\vspace{2\vspacingnodes}
\includegraphics[]{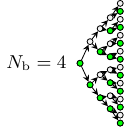}
}            \caption{Scenario tree topologies.}
            \label{fig:topologies2}     
        \end{subfigure}
        \caption{Closed-loop costs versus sample size and scenario tree topologies corresponding to different branching horizons $\bhor$
        in the deterministic setting of \Cref{exp:horizon-1}.}
        \label{fig:cost-vs-horizon-fixedsize}
    \end{figure}
    \begin{figure}
       \centering
       \includegraphics[]{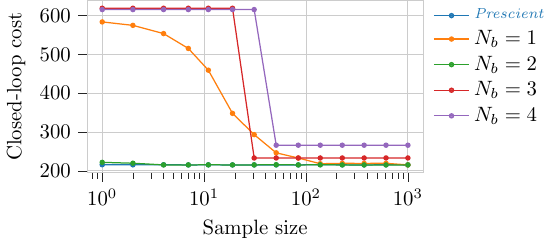}
       \caption{Cost versus sample size for the scenario tree topologies shown in \cref{fig:topologies2}, for a sample size of $\Ts=\SI{0.3}{\second}$ (compared to $\Ts=\SI{0.2}{\second}$ in \cref{fig:cost-vs-horizon-fixedsize}).}
       \label{fig:cost-vs-horizons-fixed-ts-03}
    \end{figure}
\end{experiment}

\subsubsection{Stochastic target vehicle} \label{sec:stochastic-vehicle}
Until now, the true behavior of the target vehicle was 
taken to be deterministic. In that case, there is no benefit 
to a learning-based stochastic controller. Of course, in reality, 
we cannot assume to have perfect predictions of the target 
vehicle behavior. This experiment aims to investigate the 
degree to which adding a nonzero switching probability between the 
lanes leads to hazardous situations. 

\begin{figure}[ht]
   \centering
   \includegraphics[]{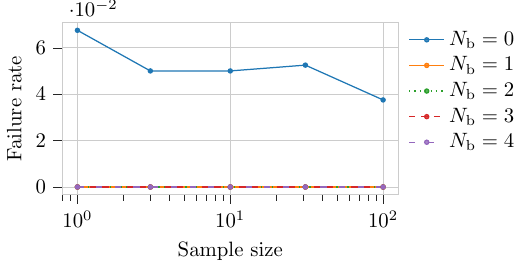}
   \caption{Fraction of infeasible problem instances over 10 random simulations of 40 time steps. Note that the lines of all nonzero branching horizon results overlap at zero.}
   \label{fig:infeasibilities-2D}
\end{figure}
\begin{experiment}[Safety comparison with deterministic MPC]

    For the sake of illustration, we model a target vehicle that switches lanes rather often by taking $\transmat = \smallmat{0.7 & 0.3 \\ 0.3 & 0.7}$. 
    We repeat the previous overtaking scenario with stochastic target 
    vehicle behavior for 10 times and record the number of times 
    the solver runs into infeasibility for each of the scenario tree 
    topologies and sample sizes. 

    \Cref{fig:infeasibilities-2D} presents the fraction of instances 
    of the optimal control problem that were detected to be infeasible.
    Since the deterministic \ac{MPC} controller does not 
    anticipate any of the lane changes, it overconfidently attempts to 
    perform an overtaking maneuver in most simulations, leading to 
    many dangerous situations, and infeasibility in approximately 5$\%$ of the simulated time steps, on average. By contrast, the 
    learning distributionally robust \ac{MPC} controllers more 
    accurately assess the situation and do not run into infeasibilities. 
    An example of a trajectory leading up to infeasibility of the deterministic controller is given in \cref{fig:example-infeas-traj}. 
    
    The random seed for all simulations was kept constant so that 
    one would expect the failure rate of the deterministic controller 
    to be independent of the sample size. The apparent dependence of the results for the deterministic controller on the sample size in \cref{fig:example-infeas-traj} are simply due to numerical errors.
    Indeed, visual inspection of the trajectories confirms that in some runs, the predicted state sequence lies very close to the boundary of the feasible set, so that a very slight deviation in solutions up to this time step may make it infeasible. 
    
    Finally, it is important to note that by careful selection of a terminal constraint set, recursive feasibility can be guaranteed for the distributionally robust controller, whereas this is not possible for the deterministic variant in this situation. This was demonstrated 
    for \iac{ACC} use case described in \cite{schuurmans_learningBased_2020a}. 
    There furthermore exists a significant body of literature on the problem of constructing invariant sets in the context of motion planning and obstacle avoidance, although most are developed with slightly different use cases in mind \cite{berntorp_PositiveInvariantSets_2020,soloperto_CollisionAvoidanceUncertain_2019,kohler_NonlinearModelPredictive_2020}. Comparing such approaches 
    on their computational efficiency, conservatism etc., is however
    beyond the scope of the current work.

    \begin{figure}[ht]
        \centering
        \begin{subfigure}{0.45\textwidth}
            \centering
            \includegraphics[width=\textwidth, trim=0 4cm 0 7cm, clip]{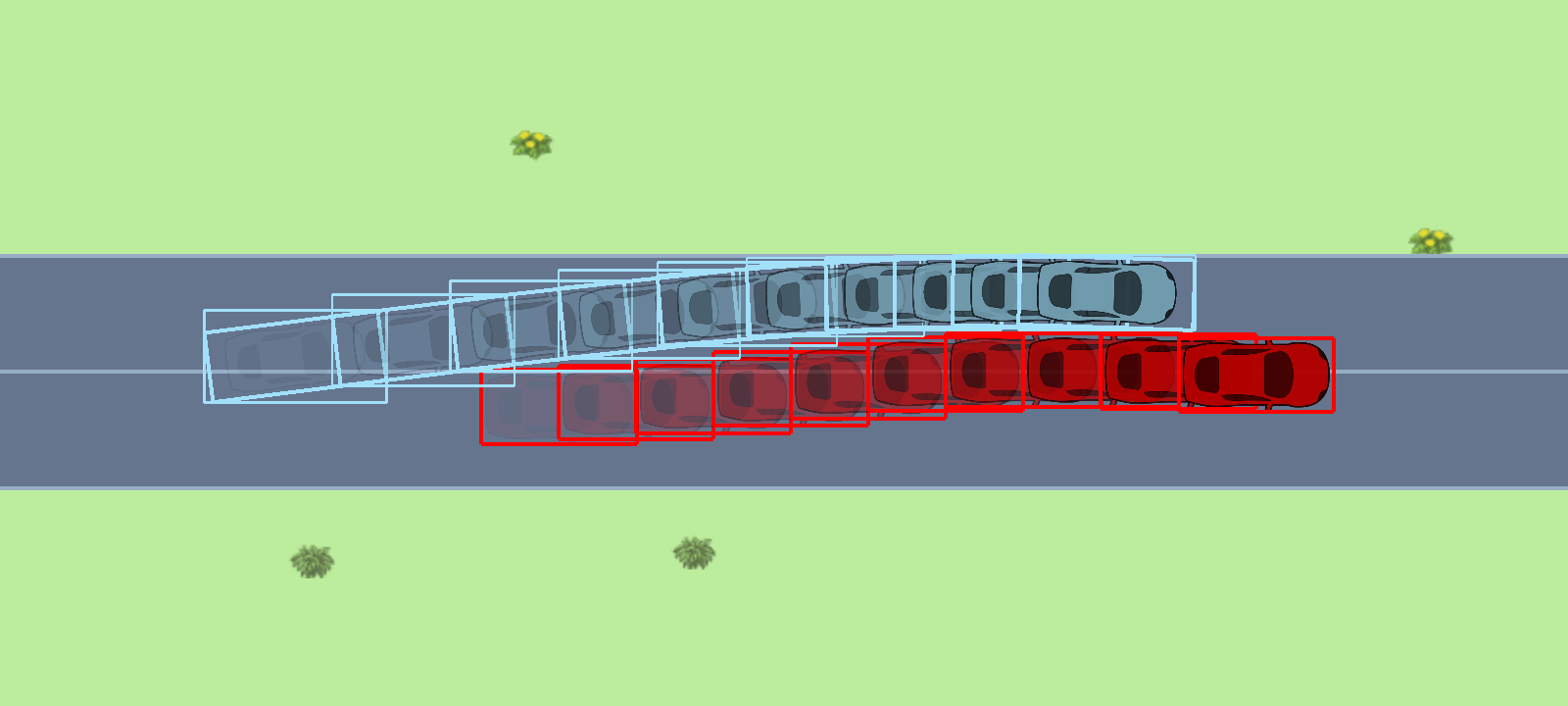}           
            
            \includegraphics[width=\textwidth]{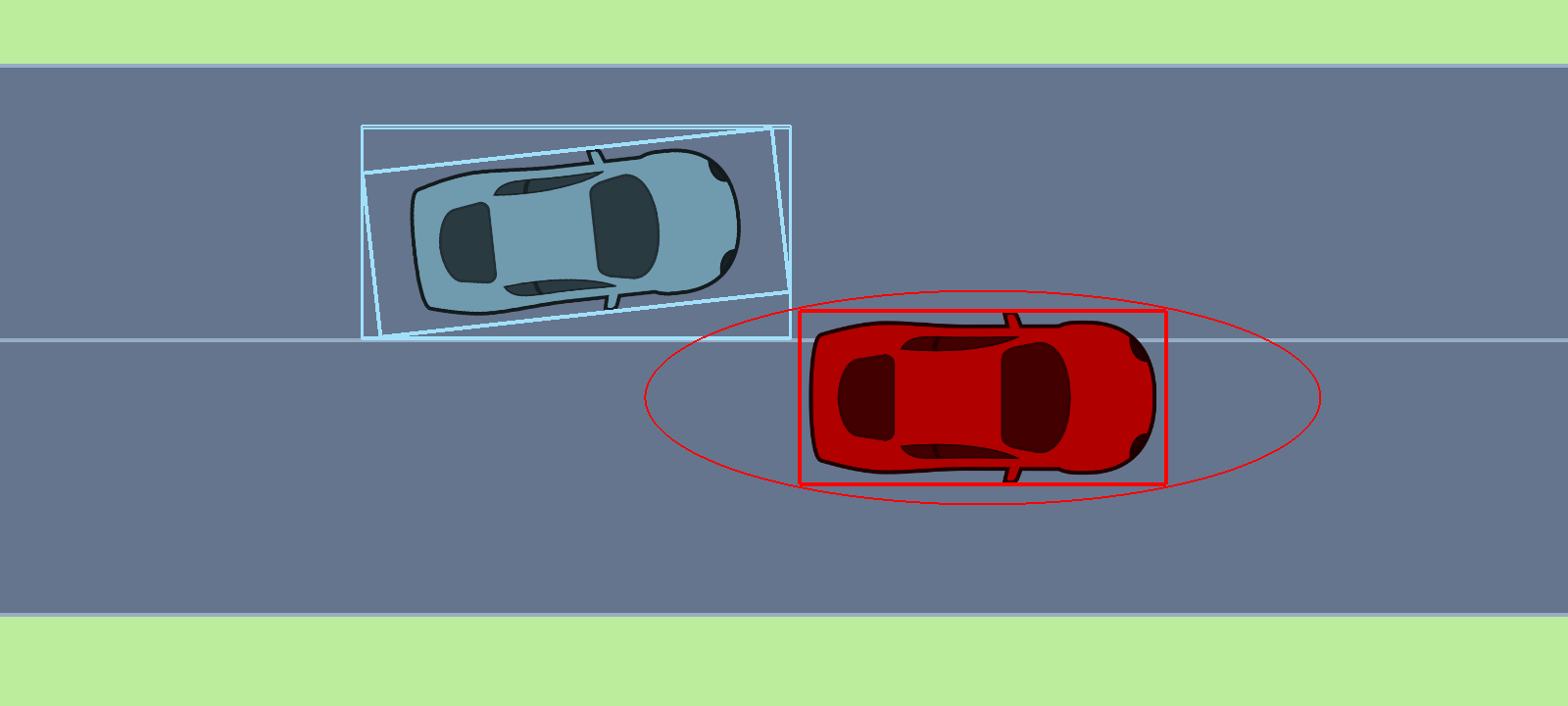}           
            
            \caption{Deterministic}
        \end{subfigure}
        \begin{subfigure}{0.45\textwidth}
            \centering
            \includegraphics[width=\textwidth, trim=0 4cm 0 7cm, clip]{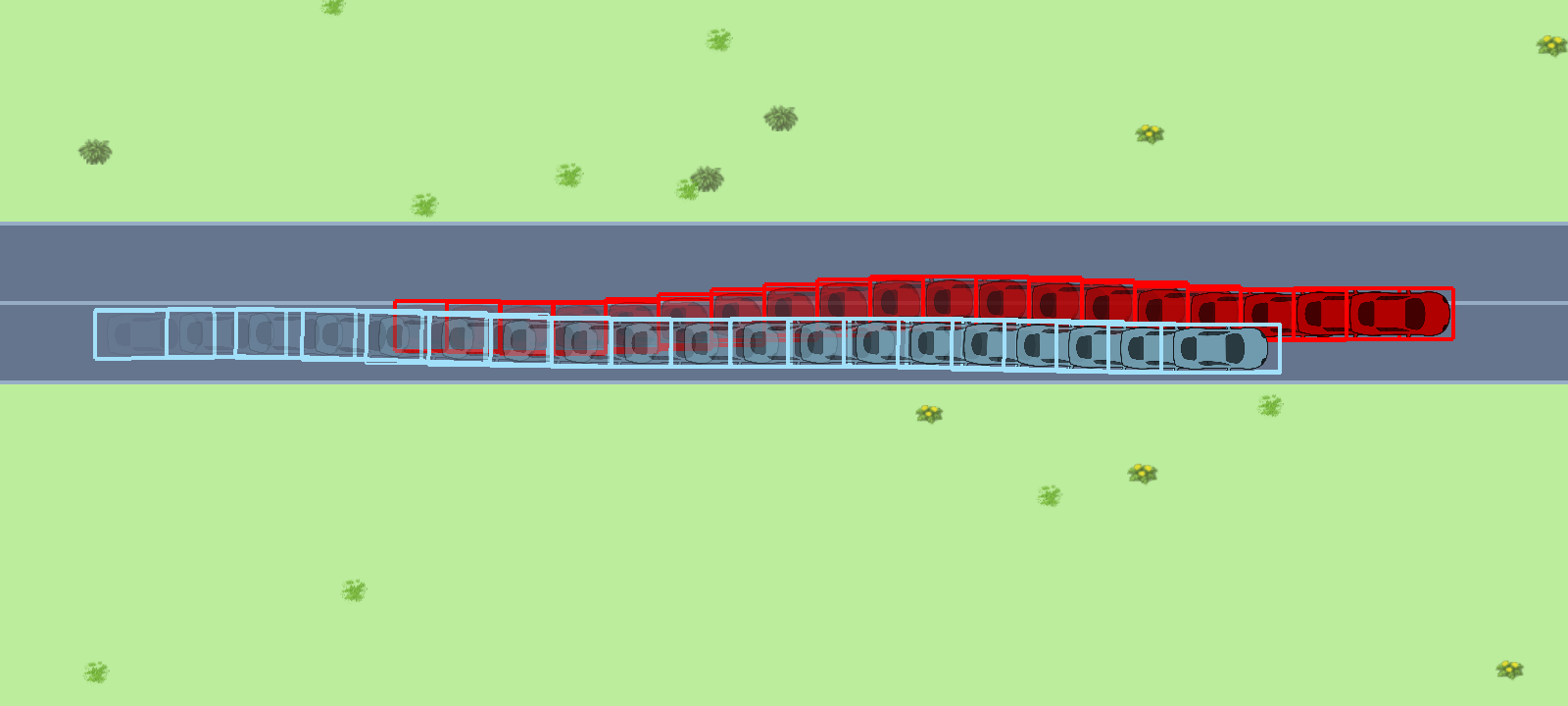}

            \includegraphics[width=\textwidth, trim=0 0cm 0 0cm, clip]{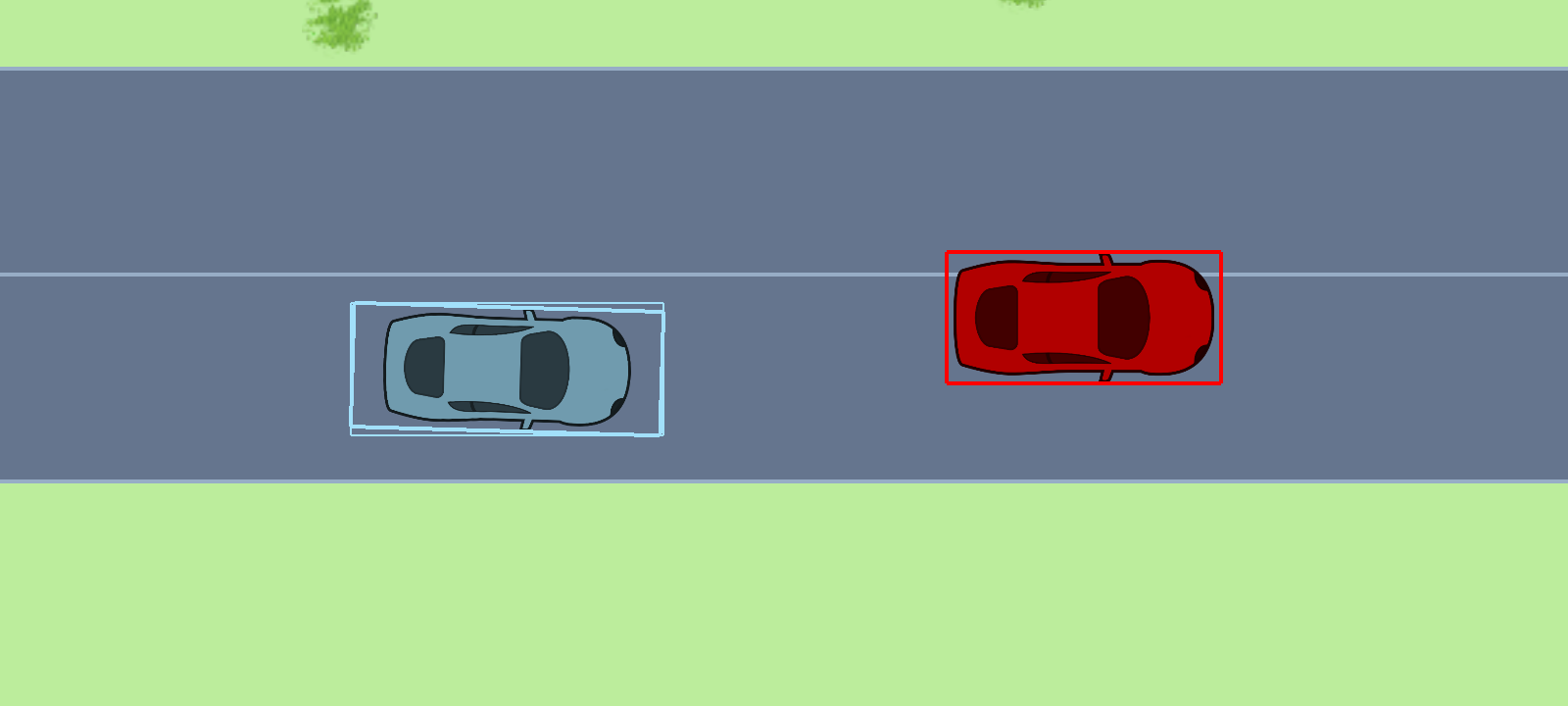}
            \caption{Distributionally robust}
        \end{subfigure}
        \caption{Snapshot of an infeasible state that occurred during the 
        simulation of the deterministic controller (a) and the 
        corresponding state (and trajectory) of the distributionally robust controller (b). The latter correctly anticipates the dangerous maneuver of the target vehicle and stays behind it. (Blue: ego vehicle; red: target vehicle.)}
        \label{fig:example-infeas-traj}
    \end{figure}
\end{experiment}

\subsection{Complexity benchmarks} \label{sec:complexity-nonconvex}

    \begin{experiment} \label{exp:benchmark-simple}
    \Cref{fig:solver-times} presents a summary of the computation 
    times measured during \Cref{exp:horizon-1,exp:horizon-2} described above. In \cref{fig:timings-increasing-size}, the runtimes are 
    shown as a function of the tree size. Analogously to the convex case \cite{schuurmans_learningBased_2020a}, 
    the complexity of the optimal control problem is linear in the 
    number of nodes in the scenario tree. Given a fixed branching horizon, it consequently grows linearly in the prediction horizon as well. 
    However, the complexity does grow exponentially in the branching horizon. As a result, the 
    choice of the branching horizon will therefore largely be dictated by 
    the available computational budget. 

    As demonstrated by \cref{fig:timings-fixed-size}, trading of the 
    prediction horizon with the branching horizon in such a way that 
    the total tree size (measured in the number of nodes) is preserved (as in \Cref{exp:horizon-2}) indeed 
    has no significant influence on the computational load. 

    \begin{figure}[ht!]
       \centering
       \begin{subfigure}[t]{0.48\columnwidth}
        \centering
           \includegraphics[]{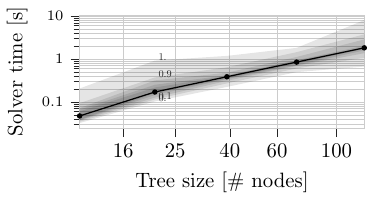}   
           \vspace{-7pt}
           \caption{Fixed prediction horizon.}
           \label{fig:timings-increasing-size}
       \end{subfigure}
       \begin{subfigure}[t]{0.48\columnwidth}
        \centering
        \includegraphics[]{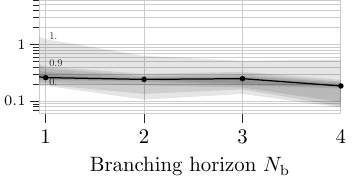}   
        \vspace{-7pt}
        \caption{Fixed tree size (approx. 25 nodes).}
        \label{fig:timings-fixed-size}
       \end{subfigure}
 
       \caption{Solver time (\IPOPT \cite{wachter_ImplementationInteriorpointFilter_2006}) versus (a) the number of nodes in the scenario tree, for a fixed prediction horizon (see \cref{fig:topologies1}); and (b) the branching horizon in the scenario tree, for a fixed number of nodes (see \cref{fig:topologies2}). The solid line depicts the median solver time over a closed-loop MPC run with the setup of \Cref{exp:horizon-1}, 
       the shaded regions (from light to dark) represent the quantiles at increments of 0.1.}
       \label{fig:solver-times}
    \end{figure}

    \end{experiment}
\revision{}{
\subsubsection{Multi-vehicle scenario} \label{sec:multi-vehicle-experiment}
\begin{figure}[ht]
   \centering
   \includegraphics[width=\textwidth]{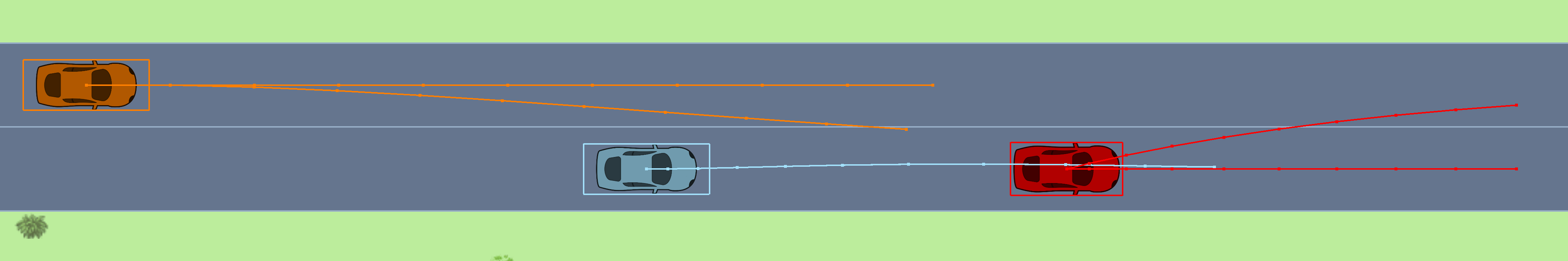}
   \caption{Scene layout for the experiment in \cref{sec:multi-vehicle-experiment}, including 2 sample trajectories for each vehicle.
   (Blue: ego vehicle, red/orange: target vehicles.)}
   \label{fig:multi-vehicle}
\end{figure}
In order to assess the capabilities of the proposed control strategy in a more complex scenario, 
we consider the set-up illustrated in \cref{fig:multi-vehicle}, involving multiple target vehicles,
and furthermore featuring a more sophisticated target vehicle driving model. 
}
\revision{}{
In this set-up, we assume, as before, that the front vehicle (shown in red) behaves according to the model 
introduced in \eqref{eq:policy-tv-simple}. Essentially, this model assumes that interaction with the 
ego vehicle is negligible. This is reasonable since the ego vehicle is driving behind the target vehicle. 

For a target vehicle approaching from the rear (shown in orange), this assumption is less easily 
justified. Instead, we consider the longitudinal control of this vehicle to be given by a
slightly modified version of the \ac{IDM} \cite{treiber_CongestedTrafficStates_2000}, which is a very commonly used 
model for simulating longitudinal driving behavior on highways. Specifically, 
assigning the index $\irear$ to the rear vehicle, we define its 
driving policy as 
\begin{equation} \label{eq:policy-interactive}
    \id{\pi}{\irear}\left( \xb, \md \right) = \smallmat{
        a_x(\xb, \md) \\ 
        \ky \left( \sub{y_\textrm{lane}}{\md} - \id{\pos_y}{\irear} \right) - \kvy \id{\vel}{\irear}_y
    }, 
\end{equation}
where the lateral component is identical to \eqref{eq:policy-tv-simple}, and the 
longitudinal control is given by 
the following model\footnote{see \Cref{tab:idm-params} for a description and numerical values for the parameters in the \ac{IDM} policy.}
\begin{equation} \label{eq:idm-model}
    a_x(\xb, \md) = \amax \tanh \left( \frac{a_0}{\amax} \left(
         1 - \left(
            \frac{v}{\sub{{\vref}}{,\md}}
        \right)^\gamma
            - 
         \varphi\left( \id{\pos}{0}_y - \id{\pos}{\irear}_y \right)  
            \left(\frac{s^\star\left( \xb \right) }{ s\left( \xb  \right) }\right)^2
    \right)
    \right), 
\end{equation}
where 
\begin{equation*}
    s\left( \xb \right) \dfn \id{\pos_{x}}{0} - \id{\pos_{x}}{\irear} \text{ and }  
        s^\star(\xb) \dfn s_0 + \id{v_{x}}{\irear} T + \frac{\id{v_{x}}{\irear} \left(\id{v_{x}}{\irear} - \id{v}{0}  \right) }{2 \sqrt{a b}}, 
\end{equation*}
are adapted closely from \ac{IDM}. The functions 
\( 
    \varphi\left( z \right) \dfn \exp\left( - z^2 \right)
\) and 
\( 
    \overline{a} \tanh \left( \frac{\argdot}{\overline{a}} \right)
\) 
were respectively introduced to
\begin{inlinelist*}
\item downweigh the interaction term between vehicle $\irear$ and the ego vehicle based on their lateral distance
\item ensure that the target vehicle accelerations remain bounded to some physically meaningful interval $[-\overline{a}, \overline{a}]$.
\end{inlinelist*}

Note that \eqref{eq:idm-model} introduces interaction between the stochastic forecasts of the rear target vehicle and 
the ego vehicle, which is a desirable property of a high-level motion planning system,
as a fully exogenous target vehicle predictions tend to result in unreasonably conservative behavior \cite{trautman_UnfreezingRobotNavigation_2010}.
}

\revision{}{
\begin{experiment} \label{exp:benchmark-multivehicle}
    \Cref{tab:solver-times-multi-vehicle} compares the solver times for 
    several closed-loop simulations of 30 time steps\footnote{We refer to the online 
    supplementary material \cite{project_page} for videos of the closed-loop trajectories.}
    of the multi-vehicle setup illustrated in \cref{fig:multi-vehicle}
    for different scenario tree structures.
    We compare 
    \begin{inlinelist*}
        \item a \textit{full uncertainty} model, where both target vehicles 
        are modeled using uncertain lane selections, resulting in a Markov chain with 
        $\nModes = 4$ modes (2 lanes per vehicle)
        \item a \textit{simplified} model, where the rear vehicle is assumed to remain 
        in the leftmost lane, and only uncertain lane change behavior is modeled 
        for the frontal target vehicle. This results in a scenario tree with $\nModes = 2$
        modes, as we had in the previous experiments.
    \end{inlinelist*}

    Besides the potentially larger number of modes, remaining sources of additional 
    complexity in the current set-up are the increased state dimension $\ns$ (12 instead of 8), 
    the nonlinearity and interaction in the rear target vehicle policy (cf. \eqref{eq:policy-interactive})
    as compared to the exogenous, affine policy \eqref{eq:policy-tv-simple} used in 
    previous experiments.

    When comparing \Cref{tab:solver-times-multi-vehicle}
    with \Cref{fig:timings-increasing-size}, we find that despite these additional complexities, 
    the solver time for a comparably sized scenario tree is not much different from
    in the simpler single-target vehicle case considered in \Cref{exp:benchmark-simple}, 
    and as we concluded before, the most important factor in determining 
    the computation time seems to be the size of the scenario tree.
    
    However, this indicates that when considering a full-uncertainty model for multiple vehicles, 
    only a relatively short prediction (or branching) horizon can be afforded to obtain an acceptable 
    computational cost. 
    In real-life practical implementations of this control method, 
    an appropriate trade-off between robustness and computation time can thus be made by 
    the designer, through the selection \begin{inlinelist*}
    \item the number of modes for each vehicle
    ---and thus, the branching factor of the scenario tree
    \item the number of branching steps $\bhor$
    \item the total horizon length $\hor$
    \item the sampling rate $\Ts$
    \end{inlinelist*}. Furthermore, it is reasonable to 
    assume that in practice, a higher level planning module
    would be used to select a small number (1 or 2) of target 
    vehicles to include in building the scenario tree, while 
    the remaining road users in the environment are taken into 
    account using a simpler model. We consider 
    a more detailed study of such a hierarchical design for future work. 
\end{experiment}
}

\begin{table}[ht] 
    \centering
    \revision{}{
    \caption{Median ($\pm$ std. dev.) solver times for \cref{exp:benchmark-multivehicle}.}
    \label{tab:solver-times-multi-vehicle}
    \begin{tabular}{lccc}
    \toprule 
                                                            & \begin{tabular}[c]{@{}l@{}}Tree size\\ $(\hor, \bhor)$\end{tabular} & \#nodes & Solver time [$\SI{}{\milli\second}$] \\
    \midrule
    \multirow{3}{*}{Simplified ($\nModes = 2$)}             & (5,1)                    & 11      & 38.17 ($\pm$ 12.70) \\
                                                            & (5, 2)                   & 19      & 70.78 ($\pm$ 26.61) \\
                                                            & (7, 2)                   & 27      & 124.47 ($\pm$ 77.25) \\
                                                        \midrule
    \multirow{3}{*}{Full uncertainty ($\nModes = 4$)}       & (5, 1)                   & 21      & 66.23 ($\pm$ 18.05)\\
                                                            & (5, 2)                   & 69      & 271.05 ($\pm$ 110.42) \\
                                                            & (7, 2)                   & 101     & 612.68 ($\pm$ 590.17) \\                                              
    \bottomrule
    \end{tabular}
    }
\end{table}

\section{Conclusion and future work} \label{sec:conclusion}
We have presented a case study on the use of a learning, distributionally robust \ac{MPC} controller for closed-loop path planning in a highway 
driving set-up. We have highlighted the most important design parameters 
and illustrated empirically how they affect controller performance, 
safety and computational load, among which a suitable trade-off is to be 
found, based on the concrete application. 

This study has highlighted some of the desirable behavioral properties 
of the controller related to its ability to autonomously and systematically
trade-off performance with caution. Combined with its 
strong theoretical underpinning \cite{schuurmans_GeneralFrameworkLearningBased_2021}, these empirical results provide a promising foundation for future developments of 
distributionally robust \ac{MPC} methodologies for autonomous navigation and other automated driving applications.
Finally, we highlight several interesting directions for further extensions and improvements to the methodology. 

\subsection{Suggestions for future work}

We identify several possible directions for further research on 
the topic, aimed primarily at extending the utility of the 
developed methodology to more practical, non-academic settings. 
We divide these directions into the following categories. 

\subsubsection{Computational}
\paragraph{Custom (parallelized) solvers} As demonstrated in the case studies, a significant bottleneck in the practicality of distributionally robust MPC techniques over scenario trees remains the computational load. 
However, in this work, only an off-the-shelf, general-purpose solver has been used, without any exploitation of the underlying scenario tree structure. For instance, the increasingly widespread availability of 
GPUs, even in embedded applications, makes an appealing case for a massively parallelized approach towards solving optimal control 
problems over the individual scenarios within a tree. (See, e.g, \cite{sampathirao_DistributedSolutionStochastic_2015,sampathirao_MassivelyParallelizableProximal_2021}.)

\subsubsection{Methodological}
\paragraph{Observers and parameter estimation} In the case where 
observations are made from continuous data (state or output measurements), 
with or without additive noise, an important question is how to determine the most likely mode to explain such an observation \cite{schuurmans_DatadrivenDistributionallyRobust_2021}.
This is particularly challenging in the case of nonlinear dynamics and additive measurement noise. 
Furthermore, it would be interesting to investigate how the misestimation at this level would translate into possible system-theoretic guarantees of the closed loop.

\paragraph{Time-varying distributions} we have considered 
static Markov chains to govern the stochastic processes. However, 
in real driving scenarios, the underlying distributions would 
likely be time-varying. In this case, the derived ambiguity bounds are 
no longer valid. However, empirical bounds based on cross-validation 
or bootstrapping could be considered, in order to obtain
online estimates of the uncertainty, which could track relatively slow-varying distributions. A trade-off would then likely arise 
between the time window/forgetting factor in the learning scheme and 
the expected rate of change of the distributions. If the distribution 
is rapidly changing, the time window in which the observations are valid 
is expected to be short, which will likely lead to rather large 
uncertainty bounds in steady state. By contrast, for a slowly changing 
distribution, a long window of observations can be gathered, and 
therefore, the estimates are expected to converge to a closer estimate of the true distribution.

\bibliographystyle{IEEEtran}
\bibliography{references}

\begin{appendix}
\crefalias{section}{appendix}\label{appendix:appdx}

    \section{Computation of smooth approximations} \label{app:smooth}
        \theoremstyle{definition}
        \newtheorem*{applemma}{Lemma}
        \begin{applemma}
            Let $\tilde{\phi}_{k}: \Re \to \Re$ be defined as 
            in \eqref{eq:smooth-abs}. Then, 
            $\tilde{\phi}_k(x) \geq |x|$ for all $k>0$ and $x \in \Re$. 
        \end{applemma}
        \begin{proof}
            We need to show that $b_k \geq \Delta(x) \dfn |x| -x \tanh{(kx)}, \forall x \in \Re$.
        Since for $k\geq 0$, both $|x|$ and $x \tanh(k x)$ are symmetric, so is $\Delta(x)$. Therefore, $\sup_{x \in \Re} \Delta(x) = \sup_{x \in \Re_+} \Delta(x)$, i.e., we may assume that $x \geq 0$ and so $|x| = x$.
        Furthermore, since for all $x \in \Re$, $x - x\tanh(kx) \leq |x| - x \tanh(kx)$, the maximum of the left-hand side will be attained for nonnegative $x$. The constraint $x \geq 0$ is therefore redundant. Thus, it suffices to show that
        \(
                b_k \geq \sup_{x} x - x \tanh(k x) .
        \)
        Let $x$ be a stationary point of $x \mapsto x - x \tanh(k x)$, then
        \[
            \begin{aligned}
                 && 1 - \tanh(kx) - 
                 \tfrac{k x}{\cosh(k x)} &= 0\\
                \Leftrightarrow&& \cosh(kx) - \sinh(kx) &= kx \\
                \Leftrightarrow&& x &= \tfrac{\Wlamb(1)}{k}.
            \end{aligned}
        \]
        It is easy to show that the unique stationary point $\tfrac{\Wlamb(1)}{k}$ is indeed a local maximizer. Since, furthermore, $x - x \tanh(kx)$ is smooth and bounded above, it is also the global maximizer. Since $b_k \geq \tfrac{\Wlamb(1)}{k}$ by construction, 
        it holds that $x \tanh(kx) + b_k \geq |x|$.
        \end{proof}

\begin{figure}[htb!]
            \centering
            \includegraphics[width=0.7\columnwidth]{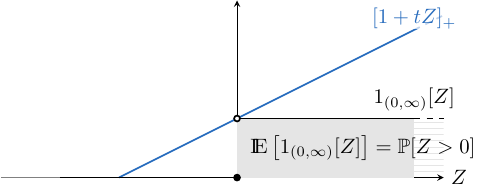}
            \caption{Graphical interpretation of the $\AVAR$ risk measure as a convex over-approximation for chance constraints.
            \revision{}{The gray area represents $\prob[Z > 0]$,
            which can be equivalently written as the expected value of $1_{\left(0,\infty\right)[Z]}$, 
            drawn in black. Drawn in blue is the convex overapproximation $[1 + t Z]_+$ for some $t \in \Re_+$.
            By minimizing over $t$, the tightest convex overapproximation is obtained \cite{shapiro_LecturesStochasticProgramming_2009,nemirovski_SafeTractableApproximations_2012}, 
            which leads to the definition of the average value-at-risk.}
            }
            \label{fig:avar-intuition}
        \end{figure}

    \revision{}{
    \section{Background on ambiguous chance constraints} \label{app:background-chanceconstraints}
        This section provides background on 
        the usage of the average-value as surrogates for chance constraints. 
        For more details, we refer the reader to \cite{nemirovski_SafeTractableApproximations_2012,shapiro_LecturesStochasticProgramming_2009}. 

        Suppose that nominally (under full knowledge of the distribution $p$),
        we would impose chance constraints of the form 
        $\prob_p[Z > 0] \leq \alpha$.
        We can equivalently denote this by 
        $\E_p \left[1_{(0, \infty)}\left( Z \right)\right] \leq \alpha$,
        as illustrated in \cref{fig:avar-intuition}.
        Since this function is nonconvex and discontinuous at zero,
        it is often approximated by its smallest convex upper bound. 
        It is easy to see that for any $t > 0$, we have 
        $\left[1+t Z \right]_{+} \geq 1_{(0, \infty)}\left( Z \right)$, 
        which holds with equality for $Z = 1$. Therefore, 
        \[ 
            \alpha \geq \min_{t > 0} \E_p \left[\left[1+t Z \right]_{+}\right] \geq \E_p \left[ 1_{(0, \infty)}( Z ) \right].
        \]
        With some straightforward manipulations, this can be reformulated as 
        \[ 
            \begin{aligned}
                \min_{t > 0} \E_p \left[\left[1+t Z \right]_{+}\right] &\leq \alpha \\
                \iff \min_{t > 0} \tfrac{1}{\alpha} \E_p \left[\left[1 + t Z \right]_{+}\right] - 1 &\leq 0\\
                \iff \min_{t > 0} \tfrac{1}{\alpha} \E_p \left[\left[t^{-1} + Z \right]_{+}\right] - t^{-1} &\leq 0\\
                \iff \min_{\gamma < 0} \tfrac{1}{\alpha} \E_p \left[\left[Z - \gamma \right]_{+}\right] + \gamma &\leq 0 
                    & \text{(Change of variables $t = -\gamma^{-1}$)} \\
                \iff \AVAR_{p}^\alpha [Z] &\leq 0, 
            \end{aligned}
        \]
        where the last equivalence follows from the fact that 
        $\AVAR_{p}^\alpha[Z] = \min_{\gamma \in \Re} \tfrac{1}{\alpha} \E_p \left[\left[Z - \gamma \right]_{+}\right] + \gamma \leq 0$
        implies that $\gamma \leq 0$, so the negativity constraint on $\gamma$ is inactive \cite[\S 6.2.4]{shapiro_LecturesStochasticProgramming_2009}.
        
        }

    \section{Detailed simulator settings} \label{app:settings}
    
        \begin{table}[H]
            \centering
            \caption{Default settings for the scenario}
            \label{tab:settings-scenario}
            \begin{tabular}{@{}ll@{}}
                \toprule
                \textbf{Setting} &\textbf{ Value} \\
                \midrule
                Sample time $\Ts$ [$\SI{}{\second}$] & 0.2 \\ 
                Transition matrix $\transmat$ & $\smallmat{0.99 & 0.01 \\ 0.05 & 0.95}$ \\ 
                Ego vehicle length [$\SI{}{\meter}$] & 4.5 \\ 
                Ego vehicle width [$\SI{}{\meter}$] & 1.8 \\ 
                Target vehicle lengths [$\SI{}{\meter}$] & $4$ \\ 
                Target vehicle widths [$\SI{}{\meter}$] & $1.9$ \\ 
                Max. velocity $\vmax$ [$\SI{}{\meter/\second}$] & 40.0 \\
                Horizon length $\hor$ & 10 \\ 
                Branching horizon $\bhor$ & 3 \\ 
                Ellipse elongation $\gamma$ & 7 \\ 
                (Nominal) violation rate $\alpha$ & 0.05 \\
                confidence parameter $\beta$ & 0.05 \\
                $\ky, \kvx, \kvy$ & 1.65, 1.83, 2.62\\
                \bottomrule
            \end{tabular}
        \end{table}
        \begin{table}[H]
            \centering
            \caption{Parameter values for the IDM model (see \cite{idm_web})}
            \label{tab:idm-params}
            \revision{}{
                {
            \begin{tabular}{lll}
            \toprule
            \textbf{Setting}& \textbf{Description} & \textbf{Value}\\
            \midrule
            $T$ & Time headway with respect to the leading vehicle [$\SI{}{\second}$]  & 0.5 \\ 
            $s_0$ & Desired spatial distance with the leading vehicle [$\SI{}{\meter}$] & 0.5 \\ 
            $a_0$ & Idle acceleration [$\SI{}{\meter/\second^2}$] & 0.3\\
            $b$ & Braking deceleration (absolute value) [$\SI{}{\meter/\second^2}$] & 0.5\\
            $\gamma$ & exponent of the free road term & 4 \\ 
            $\overline{a}$ & Physical limits on acceleration/deceleration [$\SI{}{\meter/\second^2}$] & 5\\ 
            
            \bottomrule
            \end{tabular}
            }
            }

        \end{table}

        \begin{table}[H]
            \centering   
            \caption{Default settings for the optimal control problem}
            \label{tab:settings-ctrl}
            \revision{
            \begin{tabular}{@{}lll@{}}
                \toprule
                & \multicolumn{2}{@{}c@{}}{\textbf{Value}} \\
                \cmidrule{2-3}
                \textbf{Setting} & Double integrator model & Bicycle model \\
                \midrule
                Min. controls $\underline{u}$ &
                 $\trans{
                    \smallmat{\SI{-6.5}{\meter\per\second^2} 
                    &
                    \SI{-3}{\meter\per\second^2}}
                }$ &
                $
                \trans{
                    \smallmat{
                        \SI{-6.4}{\meter\per\second^2}
                        & \SI{-3}{\degree}
                    }
                }$
                \\ 
                Max. controls $\overline{u}$ 
                & 
                $
                \trans{
                    \smallmat{
                        \SI{5.4}{\meter\per\second^2}
                        &
                        \SI{3}{\meter\per\second^2}
                    }
                }
                $
                &
                $
                \trans{
                    \smallmat{
                        \SI{5.4}{\meter\per\second^2}
                        &
                        \SI{3}{\degree}
                    }
                }$
                \\ 
                Stage cost state weights $Q$ & $\diag(0, 2,1,1)$ & $\diag(0, 2,100,5)$ \\ 
                Stage cost input weights $R$ & $I_2$ & $\smallmat{1&\\&10}$ \\
                Reference state $\xref$ & $\trans{\smallmat{0 & 0 & 30 & 0}}$ &
                $\trans{\smallmat{0 & 0 &0 & 30}}$\\ 
                \bottomrule
            \end{tabular}
            }{
            \begin{tabular}{ll}
            \toprule
            \textbf{Setting}& \textbf{Value}\\
            \midrule
            Min. controls $\underline{u}$ & 
                $
                \trans{
                    \smallmat{
                        \SI{-6.4}{\meter\per\second^2}
                        & \SI{-3}{\degree}
                    }
                }$
            \\ 
                Max. controls $\overline{u}$ 
                &
                $
                \trans{
                    \smallmat{
                        \SI{5.4}{\meter\per\second^2}
                        &
                        \SI{3}{\degree}
                    }
                }$
                \\ 
                Stage cost state weights $Q$ & $\diag(0, 2,100,5)$ \\ 
                Stage cost input weights $R$ & $\smallmat{1&\\&10}$ \\
                Reference state $\xref$ & $\trans{\smallmat{0 & 0 &0 & 30}}$\\ 
                \bottomrule
            \end{tabular}
            }
        \end{table}

\end{appendix}

\end{document}